\documentclass[
reprint,
superscriptaddress,
 amsmath,amssymb,
 aps,
prl,
floatfix,
]{revtex4-2}

\usepackage{array}
\usepackage{makecell}
\usepackage{graphicx}
\usepackage[version=4]{mhchem}
\usepackage{amssymb}
   \usepackage{amsmath}
\usepackage{physics}
\usepackage{amsfonts}
\usepackage{siunitx}
\usepackage[resetlabels]{multibib}
\newcites{supp}{References}
\usepackage{etoolbox}
\makeatletter
\AtEndEnvironment{thebibliography}{%
  \@ifundefined{@bibstop}{}{\csname bibitem@\@bibstop\endcsname}%
}
\makeatother
\makeatletter
\AfterEndPreamble{%
  \def\@bibsetup#1{\setlength{\topsep}{\z@}\NATx@bibsetnum{#1}}%
}
\makeatother

\makeatletter
\let\auto@bib@innerbib\@empty
\makeatother
\usepackage[hidelinks]{hyperref}
\renewcommand{\thetable}{\arabic{table}}  %

\begin{document}

\title{Direct Observation of Dipolar-Driven Anisotropic Quantum Projection Noise in a Solid-State Spin Ensemble}
\date{\today}
\author{Tasuku Ono}
\thanks{These authors contributed equally to this work}
\affiliation{Department of Physics, Massachusetts Institute of Technology, Cambridge, MA 02139, USA}
\affiliation{Harvard-MIT Center for Ultracold Atoms, Cambridge, MA 02139, USA}
\author{Weijie Wu}
\thanks{These authors contributed equally to this work}
\affiliation{Department of Physics, Harvard University, Cambridge, MA 02138, USA}
\affiliation{Harvard-MIT Center for Ultracold Atoms, Cambridge, MA 02139, USA}
\author{Haopu Yang}
\affiliation{Department of Physics, Harvard University, Cambridge, MA 02138, USA}
\affiliation{Harvard-MIT Center for Ultracold Atoms, Cambridge, MA 02139, USA}
\author{Lillian B. Hughes Wyatt}
\affiliation{Materials Department, University of California, Santa Barbara, CA 93106, USA}
\affiliation{Division of Engineering and Applied Science, California Institute of Technology, Pasadena, CA 91125, USA}
\author{Benjamin Brenner}
\affiliation{Department of Physics, Harvard University, Cambridge, MA 02138, USA}
\author{Che Liu}
\affiliation{Department of Physics, Harvard University, Cambridge, MA 02138, USA}
\affiliation{Harvard-MIT Center for Ultracold Atoms, Cambridge, MA 02139, USA}
\author{Collin Fan}
\affiliation{Department of Physics, Harvard University, Cambridge, MA 02138, USA}
\affiliation{Harvard-MIT Center for Ultracold Atoms, Cambridge, MA 02139, USA}
\author{Chris R. Laumann}
\affiliation{Department of Physics, Boston University, Boston, MA 02215, USA}
\affiliation{Department of Physics, Harvard University, Cambridge, MA 02138, USA}
\author{Jonathan N. Hall\'en}
\affiliation{Department of Physics, Harvard University, Cambridge, MA 02138, USA}
\affiliation{Harvard-MIT Center for Ultracold Atoms, Cambridge, MA 02139, USA}
\affiliation{Department of Physics, Boston University, Boston, MA 02215, USA}
\author{Emily J. Davis}
\affiliation{Department of Physics, Harvard University, Cambridge, MA 02138, USA}
\affiliation{Department of Physics, New York University, New York, New York 10003, USA}
\author{Ania C. Bleszynski Jayich}
\affiliation{Department of Physics, University of California, Santa Barbara, CA 93106, USA}
\author{Norman Y. Yao}
\affiliation{Department of Physics, Harvard University, Cambridge, MA 02138, USA}
\affiliation{Harvard-MIT Center for Ultracold Atoms, Cambridge, MA 02139, USA}

\begin{abstract}
The nitrogen-vacancy (NV) center in diamond is a prominent quantum-sensing platform.
Combining readout at the quantum projection noise limit with strong dipolar interactions promises substantial gains in sensitivity.
However, experimentally accessing this regime has remained a longstanding challenge. 
In this work, we demonstrate quantum-projection-noise-resolved readout of a strongly-interacting, two-dimensional ensemble of NV centers.
Our approach leverages repetitive readout via the NV's intrinsic $^{15}$N nuclear memory at a moderate magnetic field ($\sim 0.3$~T), improving the readout fidelities by nearly an order of magnitude.
This enables us to directly resolve the quantum projection noise of a coherent spin state and to watch the ensemble's intrinsic dipolar interactions shear this noise into an anisotropic profile.
Our results open the door to direct measurements of spin squeezing and entanglement-enhanced sensing in the solid state.
\end{abstract}

\maketitle

A tremendous amount of effort has been devoted to the discovery, characterization and optimization of solid-state spin defects for quantum sensing~\cite{Wolfowicz2021, Awschalom2018, Smith2019, Degen2017}. 
Among the myriad of such defects, the nitrogen-vacancy (NV) center in diamond remains paradigmatic as a particularly robust and versatile probe owing to its combination of long room-temperature spin coherences and efficient optical controls~\cite{Barry2020, Doherty2013, Levine2019}. 
These attributes have established the NV center as a workhorse platform for quantum sensing, enabling applications ranging from nanoscale magnetic and electric field imaging in condensed matter systems~\cite{Casola2018, Kolkowitz2015, Hsieh2019, Dolde2011} to intracellular thermometry in living cells~\cite{Kucsko2013}. %

\begin{figure}[t!]
\centering
\includegraphics[width=\columnwidth]{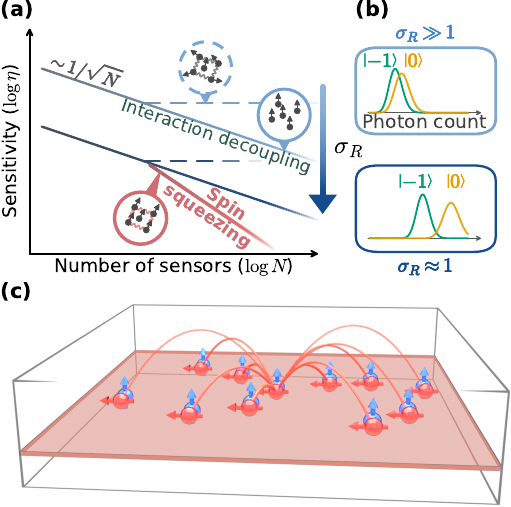}
\caption{
(a) Schematic depicting sensitivity as a function of the number of sensors in different regimes. 
The sensitivity, $\eta \sim \sigma_R/\sqrt{N\, T_2}$, can be improved by reducing the readout noise $\sigma_R$ (light blue to dark blue), by increasing the number of spins $N$, or by extending the coherence time $T_2$.
Beyond a certain density, dipolar interactions between the spins preclude reaching the standard quantum limit (dashed lines) and interaction decoupling is required to preserve the $1/\sqrt{N}$ scaling (solid lines).
At the same time, dipolar interactions can also generate spin squeezing, which yields a scaling enhancement (red line) beyond the standard quantum limit, but only for quantum-projection-noise-resolved readout ($\sigma_R \rightarrow 1$).
(b) For solid-state color centers such as the NV, the photon count distributions associated with the $\ket{0}$ and $\ket{-1}$ spin states overlap, leading to $\sigma_R \gg 1$ (top panel).
Separating these distributions improves spin-state discrimination, reducing the readout noise toward the projection-noise limit, $\sigma_R \approx 1$ (bottom panel).
(c) Our experimental system consists of a thin, delta-doped layer of NV electronic spins (red), each with an associated nitrogen nuclear spin (blue).
}
\label{fig:fig-1}
\end{figure}

Recently, the push toward further sensitivity improvements---and thus  new application domains---has centered upon two key quantities: the spin readout noise $\sigma_R$ (defined as the ratio between the total readout noise and the quantum projection noise) and 
the number of sensing spins $N$ (Fig.~\ref{fig:fig-1}a) ~\cite{Barry2020, Taylor2008, Hopper2018}.
For the former, conventional fluorescence-based readout is strongly dominated by photon shot noise (Fig.~\ref{fig:fig-1}b); in order to reach the quantum projection noise (QPN) limit, a variety of advanced readout protocols have been developed, including spin-to-charge conversion~\cite{Aslam2013, Shields2015, Jayakumar2018, Jaskula2019, Irber2021, Rovny2022} and repetitive (nuclear-spin assisted) readout~\cite{Jiang2009, Neumann2010, Steiner2010, Lovchinsky2016, Maier2026, Arunkumar2023}.
For the latter, the strategy is simply to utilize an ensemble of NV centers, leading to a sensitivity enhancement scaling as $\sim 1 / \sqrt{N}$ (the so-called standard quantum limit); for a microscopic probe of fixed size, this gain is maximized by increasing the density of NV centers, and thus the sensitivity per unit volume~\cite{Acosta2009, Wolf2015, Eichhorn2019}.

Combining these two strategies has remained a perennial challenge. 
For example, many advanced readout protocols, such as spin-to-charge conversion, were originally developed in the context of single NVs and are difficult to extend to NV ensembles.
Moreover, as the NV density grows, intrinsic magnetic dipole--dipole interactions begin to dominate the linewidth, obscuring the phase accumulated from an external signal~\cite{Bauch2020, Choi2017, Zhou2020, Davis2023, Hughes2025}.

Ironically, it is precisely this setting---dense, strongly interacting NV ensembles where readout is not photon-shot-noise limited---that promises the largest potential gains in sensing performance.
In particular, once the readout becomes QPN-limited, the very interactions that appeared detrimental can actually be harnessed to reshape the projection noise (i.e.,~spin squeezing~\cite{Kitagawa1993, Wineland1992}), enabling an entanglement-enhanced scaling of the sensitivity~\cite{Meyer2001, Gross2010, Sewell2012, Hosten2016, Block2024, Wu2025, KaplanLipkin2025}.

In this Letter, we take an important step toward the overarching goal of quantum-enhanced sensing in the solid state by demonstrating QPN-resolved readout of a dense, strongly interacting, dipolar NV ensemble.
 Our main results are twofold. 
 First, by leveraging repetitive readout in a moderate magnetic field ($B\sim 0.3$~T), we directly observe the quantum projection noise of a coherent spin state.
 We investigate four distinct NV densities and characterize the spin readout noise.
Second, we directly observe, for the first time, the interaction-driven evolution of the quantum projection noise in a dense NV ensemble.
Comparing regions with different NV densities, we find that dipolar interactions drive the emergence of anisotropic QPN: the anisotropy develops more rapidly at higher densities and is absent when the interactions are dynamically decoupled.

\emph{Resolving QPN in dense, interacting NV ensembles.---}Our system consists of a dense, strongly interacting ensemble of NV centers confined to a thin layer in bulk diamond (Fig.~\ref{fig:fig-1}c). 
As summarized in Table~\ref{tab:twisting-parameters}, we work with four distinct regions labeled A, B, C and D~\cite{SM}, each with different NV densities (controlled via electron irradiation during sample preparation). 
Each NV has an electronic spin-1 ground state, and an external magnetic field is used to lift the spin-sublevel degeneracy; we encode an effective spin-$1/2$ in the $\{\ket{m_s = 0} \equiv \ket{0}_e,\ \ket{m_s = -1} \equiv \ket{-1}_e\}$ manifold.
Conventional optical readout distinguishes these two spin states via their fluorescence under 532~nm laser illumination: $\ket{-1}_e$ appears darker than $\ket{0}_e$, owing to a non-radiative decay channel~\cite{Gruber1997, Manson2006, Doherty2013}. 
Crucially, this same channel also polarizes the NV to $\ket{0}_e$, so after only a limited number of optical cycles, the fluorescence contrast between the spin states vanishes. 
The resulting tight photon budget leaves the measurement dominated by photon shot noise, yielding $\sigma_R \gg 1$ for conventional readout in all four regions.

\begin{table}[tbp]
\centering
\setlength{\tabcolsep}{6.5pt}
\begin{tabular}{c|c|c|c|c}
\hline\hline
\makecell[c]{Spot/\\Literature}
& \makecell[c]{$N$\\(Rep-read)}
& \makecell[c]{$N$\\(XY-8)}
& \makecell[c]{$k_{\textrm{max}}$}
& \makecell[c]{$\sigma_R$} \\
\hline
A & $41 \pm 11$ & $75 \pm 23$ & $17$ & $6.8$ \\
B & $48 \pm 6$ & $36 \pm 12$ & $1.8\times10^2$ & $2.7$ \\
C & $83 \pm 20$ & $86 \pm 27$ & $45$ & $3.1$ \\
D & --- & $229 \pm 74$ & $10$ & $38$ \\
\hline\hline
\cite{Maier2026}
& \multicolumn{2}{c|}{$170$}
& $1.9\times10^4$
& $1.2$ \\
\cite{Arunkumar2023}
& \multicolumn{2}{c|}{$\sim 4\times10^8$}
& $1.1\times10^3$
& $\sim10^2$ \\
\hline\hline
\end{tabular}
\caption{Characterization of the number of spins ($N$), the maximum number of repetitive readouts $k_{\textrm{max}}$ (defined as the ratio between $T_{1n}$ and a single-shot readout time~\cite{SM}), and the readout noise $\sigma_R$, across different spots in our experimental system and the literature.
For our system, we estimate the number of NVs in two different ways: (i) by fitting the repetitive readout variance (Fig.~\ref{fig:fig-2}b) and (ii) via the $T_{2,\textrm{XY-}8}$ coherence time~\cite{SM}.
For the literature entries, $N$ is obtained as explained in the respective references.
}
\label{tab:twisting-parameters}
\end{table}

\begin{figure}[!tbp]
\centering
\includegraphics[width=\columnwidth]{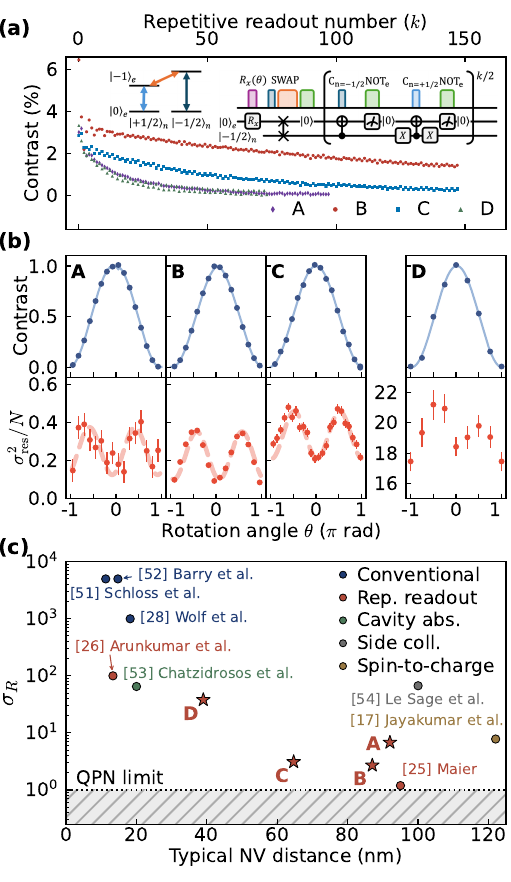}
\caption{(a) Repetitive readout contrast versus the number of readout repetitions $k$ for spots A-D. Inset depicts the repetitive readout pulse sequence.
Alternating the C$_n$NOT$_e$ gate's target nuclear spin state and subtracting the photon counts implements a differential readout that eliminates long-timescale drifts~\cite{SM}.
(b) Depicts the normalized contrast of the photon count signal $n$, which is proportional to the electronic spin polarization (top row) and the residual noise $\sigma_{\rm res}^2/N$ (bottom row) as a function of $\theta$ for all four spots. 
For spots A, B and C, the dashed line shows numerical simulations of the residual noise, accounting for hyperfine linewidth, electronic-spin initialization timescale, and nuclear-spin relaxation.
The number of spins $N$ is obtained by fitting this simulation to the residual variance. 
(c)
Readout noise $\sigma_R$ of conventional, repetitive, and other advanced readout methods in ensemble NV systems; in each case, the x-axis depicts the typical NV--NV spacing in the associated sample. Red stars correspond to our results, with parameters summarized
in Table~\ref{tab:twisting-parameters}.
}
\label{fig:fig-2}
\end{figure}

In order to improve readout toward the projection noise limit ($\sigma_R \sim 1$), we perform repetitive readout, leveraging the NV's intrinsic $^{15}$N nuclear spin
as a quantum memory~\cite{Jiang2009, Neumann2010, Steiner2010, Lovchinsky2016, Arunkumar2023, Maier2026}.
As depicted by the pulse sequence in the inset of Fig.~\ref{fig:fig-2}a, we begin by swapping the NV's electronic spin state onto the
nuclear spin, $\{\ket{m_I = -1/2} \equiv \ket{-1/2}_n,\ \ket{m_I = 1/2} \equiv \ket{+1/2}_n\}$. 
A hyperfine-selective $\pi$-pulse then implements a C$_n$NOT$_e$ gate, conditionally mapping the nuclear-spin population back onto the electron spin, which is subsequently measured via conventional spin-dependent fluorescence~\cite{Hopper2018}.
Because this measurement leaves the nuclear spin nearly unperturbed, the cycle can be repeated $k$ times, with the readout gain beginning to saturate at a characteristic repetition count $k_{\textrm{max}}$ set by the nuclear-spin lifetime $T_{1n}$.

In practice, the repeated laser illumination is actually the dominant source of nuclear-spin relaxation: hyperfine flip-flop dynamics in the electronic excited state naturally depolarize the nuclear memory.
Luckily, these flip-flop dynamics are suppressed by an external magnetic field, yielding $T_{1n} \sim B^2$~\cite{Neumann2010, Lovchinsky2016, Arunkumar2023}.
Accordingly, we extend $T_{1n}$ by applying an external magnetic field $B = 2700\,\si{G}$ using a compact permanent magnet~\footnote{The moderate field requirement here circumvents the substantial infrastructure demands of tesla-scale superconducting magnets, keeping the protocol compatible with the use of ensemble NVs as a practical sensing platform.}; this enables $k\gtrsim 100$~readout repetitions in spots B, C,
and $k\gtrsim 50$ in spots A, D (Fig.~\ref{fig:fig-2}a)~\footnote{The variation in the number of available readouts $k$ among spots is mainly due to the optimal laser power, which itself depends on spot-specific parameters such the density (see~\cite{SM}). In particular, spot A exhibited a significantly faster contrast decay than spots B and C, despite their similar NV densities. We attribute this to the electronic-spin initialization time, normalized by the optimal readout time, being approximately twice as long in spot A as in spots B and C, resulting in poorer electronic-spin polarization after each readout and, consequently, faster loss of the nuclear-spin memory.}.

Let us now attempt to directly resolve the NV ensemble's quantum projection noise in each spot.
To do so, we initialize the electronic and nuclear spins in
$\ket{0}_e\ket{-1/2}_n$ and then apply a resonant microwave pulse to prepare each NV in the state
$(\cos(\theta/2)\ket{0}_e+\sin(\theta/2)\ket{-1}_e)\ket{-1/2}_n$.
We then immediately apply repetitive readout and obtain a weighted photon-count signal, $n$, whose contrast measures the electronic spin polarization $\ev{S_z}$~\footnote{The photon count signal $n = \sum_i w_i n_i$, where $w_i$ accounts for contrast decay arising from nuclear-spin depolarization and $n_i$ is the photon count from the $i$th repetitive readout.}. 
As expected, the contrast (obtained by averaging $n$ over repeated realizations) completes one full Rabi oscillation as $\theta$ is swept, via the microwave-pulse duration, from $-\pi$ to $+\pi$ (Fig.~\ref{fig:fig-2}b, top row).
On the other hand, the quantum projection noise,
$\langle (S_z)^2\rangle-\langle S_z\rangle^2$,
is expected to oscillate twice over the same range: it is minimized at both poles of the Bloch sphere and maximized for equal-superposition states on the equator.

For each value of $\theta$, we repeat the measurement many times and determine
the variance,
$\sigma_{\rm tot}^2\equiv\operatorname{Var}(n)$.
This measured total variance contains contributions from photon shot noise as well as shot-to-shot
fluctuations in the collective spin projection (which includes the quantum projection noise but also additional technical noise).
The photon counts obey Poisson statistics, allowing us to calculate the
expected photon-shot-noise contribution, $\sigma_{\rm PSN}^2$, directly from
the measured mean photon count~\cite{SM}.
Subtracting this contribution from the total measured variance yields the
residual variance, $\sigma_{\rm res}^2
\equiv \sigma_{\rm tot}^2-\sigma_{\rm PSN}^2$.
As shown in the bottom row of Fig.~\ref{fig:fig-2}b,
$\sigma_{\rm res}^2$ exhibits two oscillations as $\theta$ is varied from
$-\pi$ to $+\pi$ for all four spots.
This twice-per-Rabi-cycle dependence is precisely characteristic of quantum projection
noise~\footnote{We note that this cannot arise from the photon shot noise subtraction since $\sigma_{\rm PSN}^2$ exhibits only one oscillation per Rabi cycle.}. 

Having demonstrated the ability to experimentally resolve the quantum projection noise, we now turn to a quantitative extraction of its magnitude, $\sigma_{\rm QPN}^2$, from the measured residual variance, $\sigma_{\rm res}^2$.
In addition to the QPN, the residual variance contains effects from the finite hyperfine linewidth, the finite electronic-spin initialization timescale, and the nuclear-spin relaxation time, $T_{1n}$.
We independently characterize each of these effects (see Supplemental Material for additional details~\cite{SM}) and then perform a numerical simulation of the full repetitive readout sequence, incorporating the effects. 
The only free parameter in the simulation is the number of spins, $N$, which sets an absolute scale for the QPN, $\sigma_{\rm QPN}^2 = N\sin^2\theta/4$.
For spots A, B and C, we find that the simulations are able to quantitatively reproduce the full $\theta$-dependence of $\sigma_{\rm res}^2$ (bottom row, Fig.~\ref{fig:fig-2}b), with the extracted values of $N$ reported in Table~\ref{tab:twisting-parameters}; for spot D, the variation in the residual variance across $\theta$ is too small to yield an accurate constraint on $N$, and thus we utilize $N$ obtained from XY-8 measurements as described below.
Indeed, we perform an independent estimate of $N$ for each spot via the $T_2$ timescale under XY-8 dynamical decoupling~\cite{SM}, and find consistency with the extracted values for spots A, B and C (Table~\ref{tab:twisting-parameters}).

Figure~\ref{fig:fig-2}c shows $\sigma_R = \sigma_{\rm tot}/\sigma_{\rm QPN}$ for each spot, demonstrating a near-order-of-magnitude improvement in readout noise (compared to conventional readout) for region B, and a $\lesssim 3\times$ improvement for regions A, C and D~\cite{SM}.  
With the perspective of ultimately combining projection-noise-limited readout with strong NV-NV interactions, Fig.~\ref{fig:fig-2}c plots $\sigma_R$ for a variety of different conventional and advanced readout protocols~\cite{Schloss2018, Barry2016, Wolf2015, Maier2026, Arunkumar2023, Chatzidrosos2017, LeSage2012, Jayakumar2018}; in each case, the sample's average NV-NV spacing, which we treat as a proxy for the interaction strength, is shown on the $x$-axis. 

\emph{Direct observation of dipolar twisting dynamics.---}That our results occupy a previously unexplored and experimentally challenging regime of near-QPN-limited readout and strong NV–NV interactions (Fig.~\ref{fig:fig-2}c) offers an intriguing possibility: namely, can one directly observe the dynamics of the NV ensemble's quantum projection noise driven by its own intrinsic dipolar interactions?

\begin{figure}[tbp]
\centering
\includegraphics[width=\columnwidth]{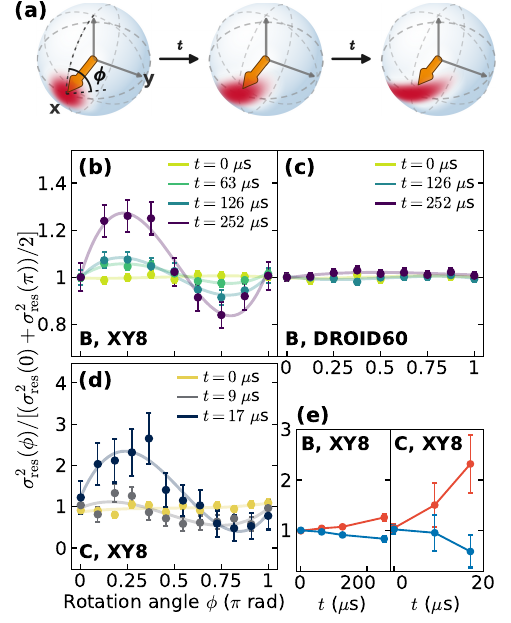}
\caption{(a) Schematic depicting the evolution of the quantum projection noise of a two-dimensional NV ensemble. The native dipolar interactions cause the projection noise to both increase in magnitude and shear. This latter effect yields an anisotropic projection noise distribution.  
(b--d) Residual noise $\sigma_{\rm res}^2(t,\phi)$ as a function of the rotation angle $\phi$ for several evolution times $t$, normalized for each $t$ by its average at $\phi = 0, \pi$. Solid lines are guides to the eye.
(b,d) For spots B and C, the dipolar interactions (under the XY-8 sequence) directly reshape the QPN. 
Under the DROID-60 decoupling sequence (spot B), the dipolar interactions are effectively SU$(2)$-symmetrized, eliminating their effect on the QPN; in this case, the noise remains isotropic, confirming that the anisotropy (observed in panels b,d) originates from  dipolar twisting dynamics~\cite{Block2024, Wu2025, KaplanLipkin2025}.
(e) Comparison between spots B and C of the rate at which anisotropy develops in the QPN as a function of dipolar evolution time, as quantified by the normalized residual noise at $\phi = \pi/4$ (red) and $3\pi/4$ (blue).
One observes that spot C, which contains a higher density of NVs and thus a stronger typical NV--NV interaction, develops QPN anisotropy significantly faster.  }
\label{fig:fig-3}
\end{figure}

Since the NV centers in our (111) sample are confined to a thin delta-doped layer ($\lesssim 7$~nm), their dipolar spin dynamics are effectively two-dimensional~\cite{Hughes2025}.
We choose to work with an NV subgroup oriented along the $[ 111 ]$-direction, whose quantization axis is perpendicular to this 2D plane, yielding the Hamiltonian \cite{Hughes2025, Wu2025},
\begin{equation}
H = -\sum_{i,j} \frac{J_0}{r_{ij}^3}\, \big( S_x^i S_x^j + S_y^i S_y^j - S_z^i S_z^j \big),
\label{eq:dipolar}
\end{equation}
where $J_0 = 52$~MHz$\cdot$nm$^3$ and $r_{ij}$ is the distance between electronic spins $i$ and $j$.
We note that for a spatially ordered array of spins (e.g.,~a square lattice), quench dynamics under $H$ are able to generate metrologically useful entanglement in the form of scalable spin squeezing~\cite{Block2024, Wu2025, KaplanLipkin2025}.
While such spin squeezing is difficult to realize in a positionally disordered ensemble such as ours, the dipolar interactions nevertheless generate so-called ``twisting dynamics,'' where the quantum projection noise of a coherent spin state becomes strongly sheared (Fig.~\ref{fig:fig-3}a). 

With this in mind, we begin by investigating region B and preparing the NVs in a coherent spin state $\ket{+}_e^{\otimes N}$ polarized along the $\hat{x}$-direction.
Next, we allow the NVs to evolve under their native dipolar interaction, while performing XY-8 dynamical decoupling to suppress extrinsic noise~\cite{Gullion1990, deLange2010, Zhou2020}.
The dipolar-driven evolution of the  quantum projection noise can be visualized through the Wigner quasiprobability distribution on the collective Bloch sphere (Fig.~\ref{fig:fig-3}a).
Under the resulting twisting dynamics, one expects the initially circular uncertainty distribution of the coherent spin state to become progressively sheared into an ellipse.

To experimentally probe this, we measure the variance of the NV ensemble's collective spin operator along an axis tilted by an angle $\phi$ from $+\hat{z}$ in the $yz$ plane, $ S_\phi^{\rm tot} = S_z^\textrm{tot} \cos\phi + S_y^\textrm{tot} \sin\phi$, where $S_\mu^\textrm{tot} =\sum_i S_\mu^i$~\cite{SM}.
As above, we subtract the photon shot noise from the total measured noise in order to obtain the residual noise, $\sigma_{\rm res}^2(t, \phi)$.
Fig.~\ref{fig:fig-3}b depicts the time evolution of the residual variance as a function of $\phi$, normalized to the average of $\phi=0$ and $\pi$.
At time $t=0$ (yellow curve, Fig.~\ref{fig:fig-3}b),  $\sigma_{\rm res}^2$ is completely independent of $\phi$, reflecting the circular nature of the coherent spin state's QPN. 
As expected from dipolar twisting dynamics, $\sigma_{\rm res}^2$ develops a clear anisotropy as a function of $\phi$ at intermediate times: the residual noise is suppressed below that of the coherent spin state along one quadrature and amplified above it along the orthogonal quadrature.  
This anisotropy increases monotonically as a function of the dipolar interaction time.
Crucially, although we measure only the residual noise, the observed anisotropy must reflect shearing of the underlying quantum projection noise; it cannot arise from the repetitive-readout process, whose noise depends only on the spin populations, and hence on $S_\phi^{\rm tot}$, which is measured to be independent of $\phi$ (see SM for details~\cite{SM}).

To isolate the role of dipolar interactions in generating the observed anisotropy, we repeat the same experiment under a different dynamical decoupling sequence, DROID-60~\cite{Choi2020, Zhou2020}.
Much like the XY-8 sequence,  DROID-60 decouples the surrounding paramagnetic spin bath; however, unlike XY-8, it also SU$(2)$-symmetrizes the NV--NV interactions, making the initial state $\ket{+}^{\otimes N}$ an eigenstate of the time evolution and arresting the dipolar dynamics. 
As strikingly illustrated in Fig.~\ref{fig:fig-3}c, the anisotropy in $\sigma_{\rm res}^2$ completely disappears  and the residual noise remains independent of $\phi$ for all times. 

Finally, we explore the twisting dynamics of the quantum projection noise in region C, which exhibits a stronger average NV--NV interaction strength than region B. 
Since the shearing of the QPN is driven by the dipolar interactions, one expects the buildup of the anisotropy in $\sigma_{\rm res}^2$ to occur on a significantly faster timescale. 
This is indeed borne out by the data (Fig.~\ref{fig:fig-3}e).
In particular, to directly track this buildup, we fix $\phi$ near the maximum and minimum of the residual noise, at \(\phi\approx\pi/4\) and \(3\pi/4\), respectively. As shown in Fig.~3(e), the separation between these two quadratures grows substantially faster for spot C than for spot B.

\emph{Conclusion and outlook.---}Our work opens the door to a number of intriguing future directions. 
First, the anisotropic reshaping of the QPN observed here is the immediate precursor to spin squeezing. 
Combining repetitive readout with techniques that suppress the effects of positional disorder could enable the direct observation of metrologically useful squeezing in dense NV ensembles~\cite{Wu2025}.
Second, QPN-resolved readout provides access to fluctuations and
correlations that are obscured by conventional fluorescence
measurements~\cite{Barry2020, Rovny2022, Maier2026}.
This capability could enable direct studies of higher-order
correlations~\cite{Schweigler2017, Pruefer2020, Schweigler2021},
non-Gaussian states~\cite{Strobel2014, HostenMagnification2016}, and
nonequilibrium many-body dynamics~\cite{ChoiDTC2017, Kucsko2018, Zu2021,
Davis2023} in strongly-interacting solid-state spin
ensembles~\cite{Rovny2024}.

\emph{Acknowledgements}---We gratefully acknowledge the insights of, and discussions with, 
Lingjie Chen, Jeff Ahlers, Winston Peloso, 
Haoyang Gao, 
Ronald Walsworth, 
Fedor Jelezko, Gerhard Wolff, Qingyun Cao,
Rouven Maier, Vadim Vorobev, and Cheng-I Ho.
This work was supported by multiple grants from the U.S. Department of Energy including BES grant DE-SC0019241 (materials synthesis and characterization) and BES grant DE-SC0026470 (numerical simulations), as well as the Army Research Office through the MURI program grant no.~W911NF-20-1-0136 (quantum dynamics experiments).
We acknowledge the use of shared facilities of the UCSB Quantum Foundry through Q-AMASE-i program (NSF DMR-1906325), the UCSB MRSEC (NSF DMR 1720256), and the Quantum Structures Facility within the UCSB California NanoSystems Institute.
T.O.\ acknowledges support from the Ezoe Memorial Recruit Foundation.
J.N.H.\ acknowledges support from The Sweden-America Foundation.
N.Y.Y.\ acknowledges support from a Brown Investigator award. 

\begingroup
\footnotesize
\raggedright
\noindent\textsuperscript{*}\,These authors contributed equally to this work.\par
\endgroup
\makeatletter
\global\let\@FMN@list\@empty
\makeatother

\bibliographystyle{apsrev4-2}
\bibliography{refs}

\begin{thebibliography}{66}%
\makeatletter
\providecommand \@ifxundefined [1]{%
 \@ifx{#1\undefined}
}%
\providecommand \@ifnum [1]{%
 \ifnum #1\expandafter \@firstoftwo
 \else \expandafter \@secondoftwo
 \fi
}%
\providecommand \@ifx [1]{%
 \ifx #1\expandafter \@firstoftwo
 \else \expandafter \@secondoftwo
 \fi
}%
\providecommand \natexlab [1]{#1}%
\providecommand \enquote  [1]{``#1''}%
\providecommand \bibnamefont  [1]{#1}%
\providecommand \bibfnamefont [1]{#1}%
\providecommand \citenamefont [1]{#1}%
\providecommand \href@noop [0]{\@secondoftwo}%
\providecommand \href [0]{\begingroup \@sanitize@url \@href}%
\providecommand \@href[1]{\@@startlink{#1}\@@href}%
\providecommand \@@href[1]{\endgroup#1\@@endlink}%
\providecommand \@sanitize@url [0]{\catcode `\\12\catcode `\$12\catcode
  `\&12\catcode `\#12\catcode `\^12\catcode `\_12\catcode `\%12\relax}%
\providecommand \@@startlink[1]{}%
\providecommand \@@endlink[0]{}%
\providecommand \url  [0]{\begingroup\@sanitize@url \@url }%
\providecommand \@url [1]{\endgroup\@href {#1}{\urlprefix }}%
\providecommand \urlprefix  [0]{URL }%
\providecommand \Eprint [0]{\href }%
\providecommand \doibase [0]{https://doi.org/}%
\providecommand \selectlanguage [0]{\@gobble}%
\providecommand \bibinfo  [0]{\@secondoftwo}%
\providecommand \bibfield  [0]{\@secondoftwo}%
\providecommand \translation [1]{[#1]}%
\providecommand \BibitemOpen [0]{}%
\providecommand \bibitemStop [0]{}%
\providecommand \bibitemNoStop [0]{.\EOS\space}%
\providecommand \EOS [0]{\spacefactor3000\relax}%
\providecommand \BibitemShut  [1]{\csname bibitem#1\endcsname}%
\let\auto@bib@innerbib\@empty
\bibitem [{\citenamefont {Wolfowicz}\ \emph {et~al.}(2021)\citenamefont
  {Wolfowicz}, \citenamefont {Heremans}, \citenamefont {Anderson},
  \citenamefont {Kanai}, \citenamefont {Seo}, \citenamefont {Gali},
  \citenamefont {Galli},\ and\ \citenamefont {Awschalom}}]{Wolfowicz2021}%
  \BibitemOpen
  \bibfield  {author} {\bibinfo {author} {\bibfnamefont {G.}~\bibnamefont
  {Wolfowicz}}, \bibinfo {author} {\bibfnamefont {F.~J.}\ \bibnamefont
  {Heremans}}, \bibinfo {author} {\bibfnamefont {C.~P.}\ \bibnamefont
  {Anderson}}, \bibinfo {author} {\bibfnamefont {S.}~\bibnamefont {Kanai}},
  \bibinfo {author} {\bibfnamefont {H.}~\bibnamefont {Seo}}, \bibinfo {author}
  {\bibfnamefont {A.}~\bibnamefont {Gali}}, \bibinfo {author} {\bibfnamefont
  {G.}~\bibnamefont {Galli}},\ and\ \bibinfo {author} {\bibfnamefont {D.~D.}\
  \bibnamefont {Awschalom}},\ }\href
  {https://doi.org/10.1038/s41578-021-00306-y} {\bibfield  {journal} {\bibinfo
  {journal} {Nature Reviews Materials}\ }\textbf {\bibinfo {volume} {6}},\
  \bibinfo {pages} {906} (\bibinfo {year} {2021})}\BibitemShut {NoStop}%
\bibitem [{\citenamefont {Awschalom}\ \emph {et~al.}(2018)\citenamefont
  {Awschalom}, \citenamefont {Hanson}, \citenamefont {Wrachtrup},\ and\
  \citenamefont {Zhou}}]{Awschalom2018}%
  \BibitemOpen
  \bibfield  {author} {\bibinfo {author} {\bibfnamefont {D.~D.}\ \bibnamefont
  {Awschalom}}, \bibinfo {author} {\bibfnamefont {R.}~\bibnamefont {Hanson}},
  \bibinfo {author} {\bibfnamefont {J.}~\bibnamefont {Wrachtrup}},\ and\
  \bibinfo {author} {\bibfnamefont {B.~B.}\ \bibnamefont {Zhou}},\ }\href
  {https://doi.org/10.1038/s41566-018-0232-2} {\bibfield  {journal} {\bibinfo
  {journal} {Nature Photonics}\ }\textbf {\bibinfo {volume} {12}},\ \bibinfo
  {pages} {516} (\bibinfo {year} {2018})}\BibitemShut {NoStop}%
\bibitem [{\citenamefont {Smith}\ \emph {et~al.}(2019)\citenamefont {Smith},
  \citenamefont {Meynell}, \citenamefont {Bleszynski~Jayich},\ and\
  \citenamefont {Meijer}}]{Smith2019}%
  \BibitemOpen
  \bibfield  {author} {\bibinfo {author} {\bibfnamefont {J.~M.}\ \bibnamefont
  {Smith}}, \bibinfo {author} {\bibfnamefont {S.~A.}\ \bibnamefont {Meynell}},
  \bibinfo {author} {\bibfnamefont {A.~C.}\ \bibnamefont {Bleszynski~Jayich}},\
  and\ \bibinfo {author} {\bibfnamefont {J.}~\bibnamefont {Meijer}},\ }\href
  {https://doi.org/10.1515/nanoph-2019-0196} {\bibfield  {journal} {\bibinfo
  {journal} {Nanophotonics}\ }\textbf {\bibinfo {volume} {8}},\ \bibinfo
  {pages} {1889} (\bibinfo {year} {2019})}\BibitemShut {NoStop}%
\bibitem [{\citenamefont {Degen}\ \emph {et~al.}(2017)\citenamefont {Degen},
  \citenamefont {Reinhard},\ and\ \citenamefont {Cappellaro}}]{Degen2017}%
  \BibitemOpen
  \bibfield  {author} {\bibinfo {author} {\bibfnamefont {C.~L.}\ \bibnamefont
  {Degen}}, \bibinfo {author} {\bibfnamefont {F.}~\bibnamefont {Reinhard}},\
  and\ \bibinfo {author} {\bibfnamefont {P.}~\bibnamefont {Cappellaro}},\
  }\href {https://doi.org/10.1103/RevModPhys.89.035002} {\bibfield  {journal}
  {\bibinfo  {journal} {Rev. Mod. Phys.}\ }\textbf {\bibinfo {volume} {89}},\
  \bibinfo {pages} {035002} (\bibinfo {year} {2017})}\BibitemShut {NoStop}%
\bibitem [{\citenamefont {Barry}\ \emph {et~al.}(2020)\citenamefont {Barry},
  \citenamefont {Schloss}, \citenamefont {Bauch}, \citenamefont {Turner},
  \citenamefont {Hart}, \citenamefont {Pham},\ and\ \citenamefont
  {Walsworth}}]{Barry2020}%
  \BibitemOpen
  \bibfield  {author} {\bibinfo {author} {\bibfnamefont {J.~F.}\ \bibnamefont
  {Barry}}, \bibinfo {author} {\bibfnamefont {J.~M.}\ \bibnamefont {Schloss}},
  \bibinfo {author} {\bibfnamefont {E.}~\bibnamefont {Bauch}}, \bibinfo
  {author} {\bibfnamefont {M.~J.}\ \bibnamefont {Turner}}, \bibinfo {author}
  {\bibfnamefont {C.~A.}\ \bibnamefont {Hart}}, \bibinfo {author}
  {\bibfnamefont {L.~M.}\ \bibnamefont {Pham}},\ and\ \bibinfo {author}
  {\bibfnamefont {R.~L.}\ \bibnamefont {Walsworth}},\ }\href
  {https://doi.org/10.1103/RevModPhys.92.015004} {\bibfield  {journal}
  {\bibinfo  {journal} {Rev. Mod. Phys.}\ }\textbf {\bibinfo {volume} {92}},\
  \bibinfo {pages} {015004} (\bibinfo {year} {2020})}\BibitemShut {NoStop}%
\bibitem [{\citenamefont {Doherty}\ \emph {et~al.}(2013)\citenamefont
  {Doherty}, \citenamefont {Manson}, \citenamefont {Delaney}, \citenamefont
  {Jelezko}, \citenamefont {Wrachtrup},\ and\ \citenamefont
  {Hollenberg}}]{Doherty2013}%
  \BibitemOpen
  \bibfield  {author} {\bibinfo {author} {\bibfnamefont {M.~W.}\ \bibnamefont
  {Doherty}}, \bibinfo {author} {\bibfnamefont {N.~B.}\ \bibnamefont {Manson}},
  \bibinfo {author} {\bibfnamefont {P.}~\bibnamefont {Delaney}}, \bibinfo
  {author} {\bibfnamefont {F.}~\bibnamefont {Jelezko}}, \bibinfo {author}
  {\bibfnamefont {J.}~\bibnamefont {Wrachtrup}},\ and\ \bibinfo {author}
  {\bibfnamefont {L.~C.~L.}\ \bibnamefont {Hollenberg}},\ }\href
  {https://doi.org/10.1016/j.physrep.2013.02.001} {\bibfield  {journal}
  {\bibinfo  {journal} {Phys. Rep.}\ }\textbf {\bibinfo {volume} {528}},\
  \bibinfo {pages} {1} (\bibinfo {year} {2013})}\BibitemShut {NoStop}%
\bibitem [{\citenamefont {Levine}\ \emph {et~al.}(2019)\citenamefont {Levine},
  \citenamefont {Turner}, \citenamefont {Kehayias}, \citenamefont {Hart},
  \citenamefont {Langellier}, \citenamefont {Trubko}, \citenamefont {Glenn},
  \citenamefont {Fu},\ and\ \citenamefont {Walsworth}}]{Levine2019}%
  \BibitemOpen
  \bibfield  {author} {\bibinfo {author} {\bibfnamefont {E.~V.}\ \bibnamefont
  {Levine}}, \bibinfo {author} {\bibfnamefont {M.~J.}\ \bibnamefont {Turner}},
  \bibinfo {author} {\bibfnamefont {P.}~\bibnamefont {Kehayias}}, \bibinfo
  {author} {\bibfnamefont {C.~A.}\ \bibnamefont {Hart}}, \bibinfo {author}
  {\bibfnamefont {N.}~\bibnamefont {Langellier}}, \bibinfo {author}
  {\bibfnamefont {R.}~\bibnamefont {Trubko}}, \bibinfo {author} {\bibfnamefont
  {D.~R.}\ \bibnamefont {Glenn}}, \bibinfo {author} {\bibfnamefont {R.~R.}\
  \bibnamefont {Fu}},\ and\ \bibinfo {author} {\bibfnamefont {R.~L.}\
  \bibnamefont {Walsworth}},\ }\href {https://doi.org/10.1515/nanoph-2019-0209}
  {\bibfield  {journal} {\bibinfo  {journal} {Nanophotonics}\ }\textbf
  {\bibinfo {volume} {8}},\ \bibinfo {pages} {1945} (\bibinfo {year}
  {2019})}\BibitemShut {NoStop}%
\bibitem [{\citenamefont {Casola}\ \emph {et~al.}(2018)\citenamefont {Casola},
  \citenamefont {van~der Sar},\ and\ \citenamefont {Yacoby}}]{Casola2018}%
  \BibitemOpen
  \bibfield  {author} {\bibinfo {author} {\bibfnamefont {F.}~\bibnamefont
  {Casola}}, \bibinfo {author} {\bibfnamefont {T.}~\bibnamefont {van~der
  Sar}},\ and\ \bibinfo {author} {\bibfnamefont {A.}~\bibnamefont {Yacoby}},\
  }\href {https://doi.org/10.1038/natrevmats.2017.88} {\bibfield  {journal}
  {\bibinfo  {journal} {Nat. Rev. Mater.}\ }\textbf {\bibinfo {volume} {3}},\
  \bibinfo {pages} {17088} (\bibinfo {year} {2018})}\BibitemShut {NoStop}%
\bibitem [{\citenamefont {Kolkowitz}\ \emph {et~al.}(2015)\citenamefont
  {Kolkowitz}, \citenamefont {Safira}, \citenamefont {High}, \citenamefont
  {Devlin}, \citenamefont {Choi}, \citenamefont {Unterreithmeier},
  \citenamefont {Patterson}, \citenamefont {Zibrov}, \citenamefont
  {Manucharyan}, \citenamefont {Park},\ and\ \citenamefont
  {Lukin}}]{Kolkowitz2015}%
  \BibitemOpen
  \bibfield  {author} {\bibinfo {author} {\bibfnamefont {S.}~\bibnamefont
  {Kolkowitz}}, \bibinfo {author} {\bibfnamefont {A.}~\bibnamefont {Safira}},
  \bibinfo {author} {\bibfnamefont {A.~A.}\ \bibnamefont {High}}, \bibinfo
  {author} {\bibfnamefont {R.~C.}\ \bibnamefont {Devlin}}, \bibinfo {author}
  {\bibfnamefont {S.}~\bibnamefont {Choi}}, \bibinfo {author} {\bibfnamefont
  {Q.~P.}\ \bibnamefont {Unterreithmeier}}, \bibinfo {author} {\bibfnamefont
  {D.}~\bibnamefont {Patterson}}, \bibinfo {author} {\bibfnamefont {A.~S.}\
  \bibnamefont {Zibrov}}, \bibinfo {author} {\bibfnamefont {V.~E.}\
  \bibnamefont {Manucharyan}}, \bibinfo {author} {\bibfnamefont
  {H.}~\bibnamefont {Park}},\ and\ \bibinfo {author} {\bibfnamefont {M.~D.}\
  \bibnamefont {Lukin}},\ }\href {https://doi.org/10.1126/science.aaa4298}
  {\bibfield  {journal} {\bibinfo  {journal} {Science}\ }\textbf {\bibinfo
  {volume} {347}},\ \bibinfo {pages} {1129} (\bibinfo {year}
  {2015})}\BibitemShut {NoStop}%
\bibitem [{\citenamefont {Hsieh}\ \emph {et~al.}(2019)\citenamefont {Hsieh},
  \citenamefont {Bhattacharyya}, \citenamefont {Zu}, \citenamefont {Mittiga},
  \citenamefont {Smart}, \citenamefont {Machado}, \citenamefont {Kobrin},
  \citenamefont {H{\"o}hn}, \citenamefont {Rui}, \citenamefont {Kamrani},
  \citenamefont {Chatterjee}, \citenamefont {Choi}, \citenamefont {Zaletel},
  \citenamefont {Struzhkin}, \citenamefont {Moore}, \citenamefont {Levitas},
  \citenamefont {Jeanloz},\ and\ \citenamefont {Yao}}]{Hsieh2019}%
  \BibitemOpen
  \bibfield  {author} {\bibinfo {author} {\bibfnamefont {S.}~\bibnamefont
  {Hsieh}}, \bibinfo {author} {\bibfnamefont {P.}~\bibnamefont
  {Bhattacharyya}}, \bibinfo {author} {\bibfnamefont {C.}~\bibnamefont {Zu}},
  \bibinfo {author} {\bibfnamefont {T.}~\bibnamefont {Mittiga}}, \bibinfo
  {author} {\bibfnamefont {T.~J.}\ \bibnamefont {Smart}}, \bibinfo {author}
  {\bibfnamefont {F.}~\bibnamefont {Machado}}, \bibinfo {author} {\bibfnamefont
  {B.}~\bibnamefont {Kobrin}}, \bibinfo {author} {\bibfnamefont {T.~O.}\
  \bibnamefont {H{\"o}hn}}, \bibinfo {author} {\bibfnamefont {N.~Z.}\
  \bibnamefont {Rui}}, \bibinfo {author} {\bibfnamefont {M.}~\bibnamefont
  {Kamrani}}, \bibinfo {author} {\bibfnamefont {S.}~\bibnamefont {Chatterjee}},
  \bibinfo {author} {\bibfnamefont {S.}~\bibnamefont {Choi}}, \bibinfo {author}
  {\bibfnamefont {M.}~\bibnamefont {Zaletel}}, \bibinfo {author} {\bibfnamefont
  {V.~V.}\ \bibnamefont {Struzhkin}}, \bibinfo {author} {\bibfnamefont {J.~E.}\
  \bibnamefont {Moore}}, \bibinfo {author} {\bibfnamefont {V.~I.}\ \bibnamefont
  {Levitas}}, \bibinfo {author} {\bibfnamefont {R.}~\bibnamefont {Jeanloz}},\
  and\ \bibinfo {author} {\bibfnamefont {N.~Y.}\ \bibnamefont {Yao}},\ }\href
  {https://doi.org/10.1126/science.aaw4352} {\bibfield  {journal} {\bibinfo
  {journal} {Science}\ }\textbf {\bibinfo {volume} {366}},\ \bibinfo {pages}
  {1349} (\bibinfo {year} {2019})}\BibitemShut {NoStop}%
\bibitem [{\citenamefont {Dolde}\ \emph {et~al.}(2011)\citenamefont {Dolde},
  \citenamefont {Fedder}, \citenamefont {Doherty}, \citenamefont {Nöbauer},
  \citenamefont {Rempp}, \citenamefont {Balasubramanian}, \citenamefont {Wolf},
  \citenamefont {Reinhard}, \citenamefont {Hollenberg}, \citenamefont
  {Jelezko},\ and\ \citenamefont {Wrachtrup}}]{Dolde2011}%
  \BibitemOpen
  \bibfield  {author} {\bibinfo {author} {\bibfnamefont {F.}~\bibnamefont
  {Dolde}}, \bibinfo {author} {\bibfnamefont {H.}~\bibnamefont {Fedder}},
  \bibinfo {author} {\bibfnamefont {M.~W.}\ \bibnamefont {Doherty}}, \bibinfo
  {author} {\bibfnamefont {T.}~\bibnamefont {Nöbauer}}, \bibinfo {author}
  {\bibfnamefont {F.}~\bibnamefont {Rempp}}, \bibinfo {author} {\bibfnamefont
  {G.}~\bibnamefont {Balasubramanian}}, \bibinfo {author} {\bibfnamefont
  {T.}~\bibnamefont {Wolf}}, \bibinfo {author} {\bibfnamefont {F.}~\bibnamefont
  {Reinhard}}, \bibinfo {author} {\bibfnamefont {L.~C.~L.}\ \bibnamefont
  {Hollenberg}}, \bibinfo {author} {\bibfnamefont {F.}~\bibnamefont
  {Jelezko}},\ and\ \bibinfo {author} {\bibfnamefont {J.}~\bibnamefont
  {Wrachtrup}},\ }\href {https://doi.org/10.1038/nphys1969} {\bibfield
  {journal} {\bibinfo  {journal} {Nature Physics}\ }\textbf {\bibinfo {volume}
  {7}},\ \bibinfo {pages} {459} (\bibinfo {year} {2011})}\BibitemShut {NoStop}%
\bibitem [{\citenamefont {Kucsko}\ \emph {et~al.}(2013)\citenamefont {Kucsko},
  \citenamefont {Maurer}, \citenamefont {Yao}, \citenamefont {Kubo},
  \citenamefont {Noh}, \citenamefont {Lo}, \citenamefont {Park},\ and\
  \citenamefont {Lukin}}]{Kucsko2013}%
  \BibitemOpen
  \bibfield  {author} {\bibinfo {author} {\bibfnamefont {G.}~\bibnamefont
  {Kucsko}}, \bibinfo {author} {\bibfnamefont {P.~C.}\ \bibnamefont {Maurer}},
  \bibinfo {author} {\bibfnamefont {N.~Y.}\ \bibnamefont {Yao}}, \bibinfo
  {author} {\bibfnamefont {M.}~\bibnamefont {Kubo}}, \bibinfo {author}
  {\bibfnamefont {H.~J.}\ \bibnamefont {Noh}}, \bibinfo {author} {\bibfnamefont
  {P.~K.}\ \bibnamefont {Lo}}, \bibinfo {author} {\bibfnamefont
  {H.}~\bibnamefont {Park}},\ and\ \bibinfo {author} {\bibfnamefont {M.~D.}\
  \bibnamefont {Lukin}},\ }\href {https://doi.org/10.1038/nature12373}
  {\bibfield  {journal} {\bibinfo  {journal} {Nature}\ }\textbf {\bibinfo
  {volume} {500}},\ \bibinfo {pages} {54} (\bibinfo {year} {2013})}\BibitemShut
  {NoStop}%
\bibitem [{\citenamefont {Taylor}\ \emph {et~al.}(2008)\citenamefont {Taylor},
  \citenamefont {Cappellaro}, \citenamefont {Childress}, \citenamefont {Jiang},
  \citenamefont {Budker}, \citenamefont {Hemmer}, \citenamefont {Yacoby},
  \citenamefont {Walsworth},\ and\ \citenamefont {Lukin}}]{Taylor2008}%
  \BibitemOpen
  \bibfield  {author} {\bibinfo {author} {\bibfnamefont {J.~M.}\ \bibnamefont
  {Taylor}}, \bibinfo {author} {\bibfnamefont {P.}~\bibnamefont {Cappellaro}},
  \bibinfo {author} {\bibfnamefont {L.}~\bibnamefont {Childress}}, \bibinfo
  {author} {\bibfnamefont {L.}~\bibnamefont {Jiang}}, \bibinfo {author}
  {\bibfnamefont {D.}~\bibnamefont {Budker}}, \bibinfo {author} {\bibfnamefont
  {P.~R.}\ \bibnamefont {Hemmer}}, \bibinfo {author} {\bibfnamefont
  {A.}~\bibnamefont {Yacoby}}, \bibinfo {author} {\bibfnamefont
  {R.}~\bibnamefont {Walsworth}},\ and\ \bibinfo {author} {\bibfnamefont
  {M.~D.}\ \bibnamefont {Lukin}},\ }\href {https://doi.org/10.1038/nphys1075}
  {\bibfield  {journal} {\bibinfo  {journal} {Nature Physics}\ }\textbf
  {\bibinfo {volume} {4}},\ \bibinfo {pages} {810} (\bibinfo {year}
  {2008})}\BibitemShut {NoStop}%
\bibitem [{\citenamefont {Hopper}\ \emph {et~al.}(2018)\citenamefont {Hopper},
  \citenamefont {Shulevitz},\ and\ \citenamefont {Bassett}}]{Hopper2018}%
  \BibitemOpen
  \bibfield  {author} {\bibinfo {author} {\bibfnamefont {D.~A.}\ \bibnamefont
  {Hopper}}, \bibinfo {author} {\bibfnamefont {H.~J.}\ \bibnamefont
  {Shulevitz}},\ and\ \bibinfo {author} {\bibfnamefont {L.~C.}\ \bibnamefont
  {Bassett}},\ }\bibfield  {journal} {\bibinfo  {journal} {Micromachines}\
  }\textbf {\bibinfo {volume} {9}},\ \href {https://doi.org/10.3390/mi9090437}
  {10.3390/mi9090437} (\bibinfo {year} {2018})\BibitemShut {NoStop}%
\bibitem [{\citenamefont {Aslam}\ \emph {et~al.}(2013)\citenamefont {Aslam},
  \citenamefont {Waldherr}, \citenamefont {Neumann}, \citenamefont {Jelezko},\
  and\ \citenamefont {Wrachtrup}}]{Aslam2013}%
  \BibitemOpen
  \bibfield  {author} {\bibinfo {author} {\bibfnamefont {N.}~\bibnamefont
  {Aslam}}, \bibinfo {author} {\bibfnamefont {G.}~\bibnamefont {Waldherr}},
  \bibinfo {author} {\bibfnamefont {P.}~\bibnamefont {Neumann}}, \bibinfo
  {author} {\bibfnamefont {F.}~\bibnamefont {Jelezko}},\ and\ \bibinfo {author}
  {\bibfnamefont {J.}~\bibnamefont {Wrachtrup}},\ }\href
  {https://doi.org/10.1088/1367-2630/15/1/013064} {\bibfield  {journal}
  {\bibinfo  {journal} {New Journal of Physics}\ }\textbf {\bibinfo {volume}
  {15}},\ \bibinfo {pages} {013064} (\bibinfo {year} {2013})}\BibitemShut
  {NoStop}%
\bibitem [{\citenamefont {Shields}\ \emph {et~al.}(2015)\citenamefont
  {Shields}, \citenamefont {Unterreithmeier}, \citenamefont {de~Leon},
  \citenamefont {Park},\ and\ \citenamefont {Lukin}}]{Shields2015}%
  \BibitemOpen
  \bibfield  {author} {\bibinfo {author} {\bibfnamefont {B.~J.}\ \bibnamefont
  {Shields}}, \bibinfo {author} {\bibfnamefont {Q.~P.}\ \bibnamefont
  {Unterreithmeier}}, \bibinfo {author} {\bibfnamefont {N.~P.}\ \bibnamefont
  {de~Leon}}, \bibinfo {author} {\bibfnamefont {H.}~\bibnamefont {Park}},\ and\
  \bibinfo {author} {\bibfnamefont {M.~D.}\ \bibnamefont {Lukin}},\ }\href
  {https://doi.org/10.1103/PhysRevLett.114.136402} {\bibfield  {journal}
  {\bibinfo  {journal} {Phys. Rev. Lett.}\ }\textbf {\bibinfo {volume} {114}},\
  \bibinfo {pages} {136402} (\bibinfo {year} {2015})}\BibitemShut {NoStop}%
\bibitem [{\citenamefont {Jayakumar}\ \emph {et~al.}(2018)\citenamefont
  {Jayakumar}, \citenamefont {Dhomkar}, \citenamefont {Henshaw},\ and\
  \citenamefont {Meriles}}]{Jayakumar2018}%
  \BibitemOpen
  \bibfield  {author} {\bibinfo {author} {\bibfnamefont {H.}~\bibnamefont
  {Jayakumar}}, \bibinfo {author} {\bibfnamefont {S.}~\bibnamefont {Dhomkar}},
  \bibinfo {author} {\bibfnamefont {J.}~\bibnamefont {Henshaw}},\ and\ \bibinfo
  {author} {\bibfnamefont {C.~A.}\ \bibnamefont {Meriles}},\ }\href
  {https://doi.org/10.1063/1.5040261} {\bibfield  {journal} {\bibinfo
  {journal} {Applied Physics Letters}\ }\textbf {\bibinfo {volume} {113}},\
  \bibinfo {pages} {122404} (\bibinfo {year} {2018})}\BibitemShut {NoStop}%
\bibitem [{\citenamefont {Jaskula}\ \emph {et~al.}(2019)\citenamefont
  {Jaskula}, \citenamefont {Shields}, \citenamefont {Bauch}, \citenamefont
  {Lukin}, \citenamefont {Trifonov},\ and\ \citenamefont
  {Walsworth}}]{Jaskula2019}%
  \BibitemOpen
  \bibfield  {author} {\bibinfo {author} {\bibfnamefont {J.-C.}\ \bibnamefont
  {Jaskula}}, \bibinfo {author} {\bibfnamefont {B.}~\bibnamefont {Shields}},
  \bibinfo {author} {\bibfnamefont {E.}~\bibnamefont {Bauch}}, \bibinfo
  {author} {\bibfnamefont {M.}~\bibnamefont {Lukin}}, \bibinfo {author}
  {\bibfnamefont {A.}~\bibnamefont {Trifonov}},\ and\ \bibinfo {author}
  {\bibfnamefont {R.}~\bibnamefont {Walsworth}},\ }\href
  {https://doi.org/10.1103/PhysRevApplied.11.064003} {\bibfield  {journal}
  {\bibinfo  {journal} {Phys. Rev. Appl.}\ }\textbf {\bibinfo {volume} {11}},\
  \bibinfo {pages} {064003} (\bibinfo {year} {2019})}\BibitemShut {NoStop}%
\bibitem [{\citenamefont {Irber}\ \emph {et~al.}(2021)\citenamefont {Irber},
  \citenamefont {Poggiali}, \citenamefont {Kong}, \citenamefont {Kieschnick},
  \citenamefont {L{\"u}hmann}, \citenamefont {Kwiatkowski}, \citenamefont
  {Meijer}, \citenamefont {Du}, \citenamefont {Shi},\ and\ \citenamefont
  {Reinhard}}]{Irber2021}%
  \BibitemOpen
  \bibfield  {author} {\bibinfo {author} {\bibfnamefont {D.~M.}\ \bibnamefont
  {Irber}}, \bibinfo {author} {\bibfnamefont {F.}~\bibnamefont {Poggiali}},
  \bibinfo {author} {\bibfnamefont {F.}~\bibnamefont {Kong}}, \bibinfo {author}
  {\bibfnamefont {M.}~\bibnamefont {Kieschnick}}, \bibinfo {author}
  {\bibfnamefont {T.}~\bibnamefont {L{\"u}hmann}}, \bibinfo {author}
  {\bibfnamefont {D.}~\bibnamefont {Kwiatkowski}}, \bibinfo {author}
  {\bibfnamefont {J.}~\bibnamefont {Meijer}}, \bibinfo {author} {\bibfnamefont
  {J.}~\bibnamefont {Du}}, \bibinfo {author} {\bibfnamefont {F.}~\bibnamefont
  {Shi}},\ and\ \bibinfo {author} {\bibfnamefont {F.}~\bibnamefont
  {Reinhard}},\ }\href {https://doi.org/10.1038/s41467-020-20755-3} {\bibfield
  {journal} {\bibinfo  {journal} {Nature Communications}\ }\textbf {\bibinfo
  {volume} {12}},\ \bibinfo {pages} {532} (\bibinfo {year} {2021})}\BibitemShut
  {NoStop}%
\bibitem [{\citenamefont {Rovny}\ \emph {et~al.}(2022)\citenamefont {Rovny},
  \citenamefont {Yuan}, \citenamefont {Fitzpatrick}, \citenamefont {Abdalla},
  \citenamefont {Futamura}, \citenamefont {Fox}, \citenamefont {Cambria},
  \citenamefont {Kolkowitz},\ and\ \citenamefont {de~Leon}}]{Rovny2022}%
  \BibitemOpen
  \bibfield  {author} {\bibinfo {author} {\bibfnamefont {J.}~\bibnamefont
  {Rovny}}, \bibinfo {author} {\bibfnamefont {Z.}~\bibnamefont {Yuan}},
  \bibinfo {author} {\bibfnamefont {M.}~\bibnamefont {Fitzpatrick}}, \bibinfo
  {author} {\bibfnamefont {A.~I.}\ \bibnamefont {Abdalla}}, \bibinfo {author}
  {\bibfnamefont {L.}~\bibnamefont {Futamura}}, \bibinfo {author}
  {\bibfnamefont {C.}~\bibnamefont {Fox}}, \bibinfo {author} {\bibfnamefont
  {M.~C.}\ \bibnamefont {Cambria}}, \bibinfo {author} {\bibfnamefont
  {S.}~\bibnamefont {Kolkowitz}},\ and\ \bibinfo {author} {\bibfnamefont
  {N.~P.}\ \bibnamefont {de~Leon}},\ }\href
  {https://doi.org/10.1126/science.ade9858} {\bibfield  {journal} {\bibinfo
  {journal} {Science}\ }\textbf {\bibinfo {volume} {378}},\ \bibinfo {pages}
  {1301} (\bibinfo {year} {2022})}\BibitemShut {NoStop}%
\bibitem [{\citenamefont {Jiang}\ \emph {et~al.}(2009)\citenamefont {Jiang},
  \citenamefont {Hodges}, \citenamefont {Maze}, \citenamefont {Maurer},
  \citenamefont {Taylor}, \citenamefont {Cory}, \citenamefont {Hemmer},
  \citenamefont {Walsworth}, \citenamefont {Yacoby}, \citenamefont {Zibrov},\
  and\ \citenamefont {Lukin}}]{Jiang2009}%
  \BibitemOpen
  \bibfield  {author} {\bibinfo {author} {\bibfnamefont {L.}~\bibnamefont
  {Jiang}}, \bibinfo {author} {\bibfnamefont {J.~S.}\ \bibnamefont {Hodges}},
  \bibinfo {author} {\bibfnamefont {J.~R.}\ \bibnamefont {Maze}}, \bibinfo
  {author} {\bibfnamefont {P.}~\bibnamefont {Maurer}}, \bibinfo {author}
  {\bibfnamefont {J.~M.}\ \bibnamefont {Taylor}}, \bibinfo {author}
  {\bibfnamefont {D.~G.}\ \bibnamefont {Cory}}, \bibinfo {author}
  {\bibfnamefont {P.~R.}\ \bibnamefont {Hemmer}}, \bibinfo {author}
  {\bibfnamefont {R.~L.}\ \bibnamefont {Walsworth}}, \bibinfo {author}
  {\bibfnamefont {A.}~\bibnamefont {Yacoby}}, \bibinfo {author} {\bibfnamefont
  {A.~S.}\ \bibnamefont {Zibrov}},\ and\ \bibinfo {author} {\bibfnamefont
  {M.~D.}\ \bibnamefont {Lukin}},\ }\href
  {https://doi.org/10.1126/science.1176496} {\bibfield  {journal} {\bibinfo
  {journal} {Science}\ }\textbf {\bibinfo {volume} {326}},\ \bibinfo {pages}
  {267} (\bibinfo {year} {2009})}\BibitemShut {NoStop}%
\bibitem [{\citenamefont {Neumann}\ \emph {et~al.}(2010)\citenamefont
  {Neumann}, \citenamefont {Beck}, \citenamefont {Steiner}, \citenamefont
  {Rempp}, \citenamefont {Fedder}, \citenamefont {Hemmer}, \citenamefont
  {Wrachtrup},\ and\ \citenamefont {Jelezko}}]{Neumann2010}%
  \BibitemOpen
  \bibfield  {author} {\bibinfo {author} {\bibfnamefont {P.}~\bibnamefont
  {Neumann}}, \bibinfo {author} {\bibfnamefont {J.}~\bibnamefont {Beck}},
  \bibinfo {author} {\bibfnamefont {M.}~\bibnamefont {Steiner}}, \bibinfo
  {author} {\bibfnamefont {F.}~\bibnamefont {Rempp}}, \bibinfo {author}
  {\bibfnamefont {H.}~\bibnamefont {Fedder}}, \bibinfo {author} {\bibfnamefont
  {P.~R.}\ \bibnamefont {Hemmer}}, \bibinfo {author} {\bibfnamefont
  {J.}~\bibnamefont {Wrachtrup}},\ and\ \bibinfo {author} {\bibfnamefont
  {F.}~\bibnamefont {Jelezko}},\ }\href
  {https://doi.org/10.1126/science.1189075} {\bibfield  {journal} {\bibinfo
  {journal} {Science}\ }\textbf {\bibinfo {volume} {329}},\ \bibinfo {pages}
  {542} (\bibinfo {year} {2010})}\BibitemShut {NoStop}%
\bibitem [{\citenamefont {Steiner}\ \emph {et~al.}(2010)\citenamefont
  {Steiner}, \citenamefont {Neumann}, \citenamefont {Beck}, \citenamefont
  {Jelezko},\ and\ \citenamefont {Wrachtrup}}]{Steiner2010}%
  \BibitemOpen
  \bibfield  {author} {\bibinfo {author} {\bibfnamefont {M.}~\bibnamefont
  {Steiner}}, \bibinfo {author} {\bibfnamefont {P.}~\bibnamefont {Neumann}},
  \bibinfo {author} {\bibfnamefont {J.}~\bibnamefont {Beck}}, \bibinfo {author}
  {\bibfnamefont {F.}~\bibnamefont {Jelezko}},\ and\ \bibinfo {author}
  {\bibfnamefont {J.}~\bibnamefont {Wrachtrup}},\ }\href
  {https://doi.org/10.1103/PhysRevB.81.035205} {\bibfield  {journal} {\bibinfo
  {journal} {Phys. Rev. B}\ }\textbf {\bibinfo {volume} {81}},\ \bibinfo
  {pages} {035205} (\bibinfo {year} {2010})}\BibitemShut {NoStop}%
\bibitem [{\citenamefont {Lovchinsky}\ \emph {et~al.}(2016)\citenamefont
  {Lovchinsky}, \citenamefont {Sushkov}, \citenamefont {Urbach}, \citenamefont
  {de~Leon}, \citenamefont {Choi}, \citenamefont {De~Greve}, \citenamefont
  {Evans}, \citenamefont {Gertner}, \citenamefont {Bersin}, \citenamefont
  {M{\"u}ller}, \citenamefont {McGuinness}, \citenamefont {Jelezko},
  \citenamefont {Walsworth}, \citenamefont {Park},\ and\ \citenamefont
  {Lukin}}]{Lovchinsky2016}%
  \BibitemOpen
  \bibfield  {author} {\bibinfo {author} {\bibfnamefont {I.}~\bibnamefont
  {Lovchinsky}}, \bibinfo {author} {\bibfnamefont {A.~O.}\ \bibnamefont
  {Sushkov}}, \bibinfo {author} {\bibfnamefont {E.}~\bibnamefont {Urbach}},
  \bibinfo {author} {\bibfnamefont {N.~P.}\ \bibnamefont {de~Leon}}, \bibinfo
  {author} {\bibfnamefont {S.}~\bibnamefont {Choi}}, \bibinfo {author}
  {\bibfnamefont {K.}~\bibnamefont {De~Greve}}, \bibinfo {author}
  {\bibfnamefont {R.}~\bibnamefont {Evans}}, \bibinfo {author} {\bibfnamefont
  {R.}~\bibnamefont {Gertner}}, \bibinfo {author} {\bibfnamefont
  {E.}~\bibnamefont {Bersin}}, \bibinfo {author} {\bibfnamefont
  {C.}~\bibnamefont {M{\"u}ller}}, \bibinfo {author} {\bibfnamefont
  {L.}~\bibnamefont {McGuinness}}, \bibinfo {author} {\bibfnamefont
  {F.}~\bibnamefont {Jelezko}}, \bibinfo {author} {\bibfnamefont {R.~L.}\
  \bibnamefont {Walsworth}}, \bibinfo {author} {\bibfnamefont {H.}~\bibnamefont
  {Park}},\ and\ \bibinfo {author} {\bibfnamefont {M.~D.}\ \bibnamefont
  {Lukin}},\ }\href {https://doi.org/10.1126/science.aad8022} {\bibfield
  {journal} {\bibinfo  {journal} {Science}\ }\textbf {\bibinfo {volume}
  {351}},\ \bibinfo {pages} {836} (\bibinfo {year} {2016})}\BibitemShut
  {NoStop}%
\bibitem [{\citenamefont {Maier}\ \emph {et~al.}(2026)\citenamefont {Maier},
  \citenamefont {Ho}, \citenamefont {Denisenko}, \citenamefont {Davydova},
  \citenamefont {Knittel}, \citenamefont {Wrachtrup},\ and\ \citenamefont
  {Vorobyov}}]{Maier2026}%
  \BibitemOpen
  \bibfield  {author} {\bibinfo {author} {\bibfnamefont {R.}~\bibnamefont
  {Maier}}, \bibinfo {author} {\bibfnamefont {C.-I.}\ \bibnamefont {Ho}},
  \bibinfo {author} {\bibfnamefont {A.}~\bibnamefont {Denisenko}}, \bibinfo
  {author} {\bibfnamefont {M.}~\bibnamefont {Davydova}}, \bibinfo {author}
  {\bibfnamefont {P.}~\bibnamefont {Knittel}}, \bibinfo {author} {\bibfnamefont
  {J.}~\bibnamefont {Wrachtrup}},\ and\ \bibinfo {author} {\bibfnamefont
  {V.}~\bibnamefont {Vorobyov}},\ }\href
  {https://doi.org/10.1038/s41467-026-72721-0} {\bibfield  {journal} {\bibinfo
  {journal} {Nature Communications}\ }\textbf {\bibinfo {volume} {17}},\
  \bibinfo {pages} {4028} (\bibinfo {year} {2026})}\BibitemShut {NoStop}%
\bibitem [{\citenamefont {Arunkumar}\ \emph {et~al.}(2023)\citenamefont
  {Arunkumar}, \citenamefont {Olsson}, \citenamefont {Oon}, \citenamefont
  {Hart}, \citenamefont {Bucher}, \citenamefont {Glenn}, \citenamefont {Lukin},
  \citenamefont {Park}, \citenamefont {Ham},\ and\ \citenamefont
  {Walsworth}}]{Arunkumar2023}%
  \BibitemOpen
  \bibfield  {author} {\bibinfo {author} {\bibfnamefont {N.}~\bibnamefont
  {Arunkumar}}, \bibinfo {author} {\bibfnamefont {K.~S.}\ \bibnamefont
  {Olsson}}, \bibinfo {author} {\bibfnamefont {J.~T.}\ \bibnamefont {Oon}},
  \bibinfo {author} {\bibfnamefont {C.~A.}\ \bibnamefont {Hart}}, \bibinfo
  {author} {\bibfnamefont {D.~B.}\ \bibnamefont {Bucher}}, \bibinfo {author}
  {\bibfnamefont {D.~R.}\ \bibnamefont {Glenn}}, \bibinfo {author}
  {\bibfnamefont {M.~D.}\ \bibnamefont {Lukin}}, \bibinfo {author}
  {\bibfnamefont {H.}~\bibnamefont {Park}}, \bibinfo {author} {\bibfnamefont
  {D.}~\bibnamefont {Ham}},\ and\ \bibinfo {author} {\bibfnamefont {R.~L.}\
  \bibnamefont {Walsworth}},\ }\href
  {https://doi.org/10.1103/PhysRevLett.131.100801} {\bibfield  {journal}
  {\bibinfo  {journal} {Phys. Rev. Lett.}\ }\textbf {\bibinfo {volume} {131}},\
  \bibinfo {pages} {100801} (\bibinfo {year} {2023})}\BibitemShut {NoStop}%
\bibitem [{\citenamefont {Acosta}\ \emph {et~al.}(2009)\citenamefont {Acosta},
  \citenamefont {Bauch}, \citenamefont {Ledbetter}, \citenamefont {Santori},
  \citenamefont {Fu}, \citenamefont {Barclay}, \citenamefont {Beausoleil},
  \citenamefont {Linget}, \citenamefont {Roch}, \citenamefont {Treussart},
  \citenamefont {Chemerisov}, \citenamefont {Gawlik},\ and\ \citenamefont
  {Budker}}]{Acosta2009}%
  \BibitemOpen
  \bibfield  {author} {\bibinfo {author} {\bibfnamefont {V.~M.}\ \bibnamefont
  {Acosta}}, \bibinfo {author} {\bibfnamefont {E.}~\bibnamefont {Bauch}},
  \bibinfo {author} {\bibfnamefont {M.~P.}\ \bibnamefont {Ledbetter}}, \bibinfo
  {author} {\bibfnamefont {C.}~\bibnamefont {Santori}}, \bibinfo {author}
  {\bibfnamefont {K.-M.~C.}\ \bibnamefont {Fu}}, \bibinfo {author}
  {\bibfnamefont {P.~E.}\ \bibnamefont {Barclay}}, \bibinfo {author}
  {\bibfnamefont {R.~G.}\ \bibnamefont {Beausoleil}}, \bibinfo {author}
  {\bibfnamefont {H.}~\bibnamefont {Linget}}, \bibinfo {author} {\bibfnamefont
  {J.~F.}\ \bibnamefont {Roch}}, \bibinfo {author} {\bibfnamefont
  {F.}~\bibnamefont {Treussart}}, \bibinfo {author} {\bibfnamefont
  {S.}~\bibnamefont {Chemerisov}}, \bibinfo {author} {\bibfnamefont
  {W.}~\bibnamefont {Gawlik}},\ and\ \bibinfo {author} {\bibfnamefont
  {D.}~\bibnamefont {Budker}},\ }\href
  {https://doi.org/10.1103/PhysRevB.80.115202} {\bibfield  {journal} {\bibinfo
  {journal} {Phys. Rev. B}\ }\textbf {\bibinfo {volume} {80}},\ \bibinfo
  {pages} {115202} (\bibinfo {year} {2009})}\BibitemShut {NoStop}%
\bibitem [{\citenamefont {Wolf}\ \emph {et~al.}(2015)\citenamefont {Wolf},
  \citenamefont {Neumann}, \citenamefont {Nakamura}, \citenamefont {Sumiya},
  \citenamefont {Ohshima}, \citenamefont {Isoya},\ and\ \citenamefont
  {Wrachtrup}}]{Wolf2015}%
  \BibitemOpen
  \bibfield  {author} {\bibinfo {author} {\bibfnamefont {T.}~\bibnamefont
  {Wolf}}, \bibinfo {author} {\bibfnamefont {P.}~\bibnamefont {Neumann}},
  \bibinfo {author} {\bibfnamefont {K.}~\bibnamefont {Nakamura}}, \bibinfo
  {author} {\bibfnamefont {H.}~\bibnamefont {Sumiya}}, \bibinfo {author}
  {\bibfnamefont {T.}~\bibnamefont {Ohshima}}, \bibinfo {author} {\bibfnamefont
  {J.}~\bibnamefont {Isoya}},\ and\ \bibinfo {author} {\bibfnamefont
  {J.}~\bibnamefont {Wrachtrup}},\ }\href
  {https://doi.org/10.1103/PhysRevX.5.041001} {\bibfield  {journal} {\bibinfo
  {journal} {Phys. Rev. X}\ }\textbf {\bibinfo {volume} {5}},\ \bibinfo {pages}
  {041001} (\bibinfo {year} {2015})}\BibitemShut {NoStop}%
\bibitem [{\citenamefont {Eichhorn}\ \emph {et~al.}(2019)\citenamefont
  {Eichhorn}, \citenamefont {McLellan},\ and\ \citenamefont
  {Bleszynski~Jayich}}]{Eichhorn2019}%
  \BibitemOpen
  \bibfield  {author} {\bibinfo {author} {\bibfnamefont {T.~R.}\ \bibnamefont
  {Eichhorn}}, \bibinfo {author} {\bibfnamefont {C.~A.}\ \bibnamefont
  {McLellan}},\ and\ \bibinfo {author} {\bibfnamefont {A.~C.}\ \bibnamefont
  {Bleszynski~Jayich}},\ }\href
  {https://doi.org/10.1103/PhysRevMaterials.3.113802} {\bibfield  {journal}
  {\bibinfo  {journal} {Phys. Rev. Mater.}\ }\textbf {\bibinfo {volume} {3}},\
  \bibinfo {pages} {113802} (\bibinfo {year} {2019})}\BibitemShut {NoStop}%
\bibitem [{\citenamefont {Bauch}\ \emph {et~al.}(2020)\citenamefont {Bauch},
  \citenamefont {Singh}, \citenamefont {Lee}, \citenamefont {Hart},
  \citenamefont {Schloss}, \citenamefont {Turner}, \citenamefont {Barry},
  \citenamefont {Pham}, \citenamefont {Bar-Gill}, \citenamefont {Yelin},\ and\
  \citenamefont {Walsworth}}]{Bauch2020}%
  \BibitemOpen
  \bibfield  {author} {\bibinfo {author} {\bibfnamefont {E.}~\bibnamefont
  {Bauch}}, \bibinfo {author} {\bibfnamefont {S.}~\bibnamefont {Singh}},
  \bibinfo {author} {\bibfnamefont {J.}~\bibnamefont {Lee}}, \bibinfo {author}
  {\bibfnamefont {C.~A.}\ \bibnamefont {Hart}}, \bibinfo {author}
  {\bibfnamefont {J.~M.}\ \bibnamefont {Schloss}}, \bibinfo {author}
  {\bibfnamefont {M.~J.}\ \bibnamefont {Turner}}, \bibinfo {author}
  {\bibfnamefont {J.~F.}\ \bibnamefont {Barry}}, \bibinfo {author}
  {\bibfnamefont {L.~M.}\ \bibnamefont {Pham}}, \bibinfo {author}
  {\bibfnamefont {N.}~\bibnamefont {Bar-Gill}}, \bibinfo {author}
  {\bibfnamefont {S.~F.}\ \bibnamefont {Yelin}},\ and\ \bibinfo {author}
  {\bibfnamefont {R.~L.}\ \bibnamefont {Walsworth}},\ }\href
  {https://doi.org/10.1103/PhysRevB.102.134210} {\bibfield  {journal} {\bibinfo
   {journal} {Phys. Rev. B}\ }\textbf {\bibinfo {volume} {102}},\ \bibinfo
  {pages} {134210} (\bibinfo {year} {2020})}\BibitemShut {NoStop}%
\bibitem [{\citenamefont {Choi}\ \emph
  {et~al.}(2017{\natexlab{a}})\citenamefont {Choi}, \citenamefont {Choi},
  \citenamefont {Kucsko}, \citenamefont {Maurer}, \citenamefont {Shields},
  \citenamefont {Sumiya}, \citenamefont {Onoda}, \citenamefont {Isoya},
  \citenamefont {Demler}, \citenamefont {Jelezko}, \citenamefont {Yao},\ and\
  \citenamefont {Lukin}}]{Choi2017}%
  \BibitemOpen
  \bibfield  {author} {\bibinfo {author} {\bibfnamefont {J.}~\bibnamefont
  {Choi}}, \bibinfo {author} {\bibfnamefont {S.}~\bibnamefont {Choi}}, \bibinfo
  {author} {\bibfnamefont {G.}~\bibnamefont {Kucsko}}, \bibinfo {author}
  {\bibfnamefont {P.~C.}\ \bibnamefont {Maurer}}, \bibinfo {author}
  {\bibfnamefont {B.~J.}\ \bibnamefont {Shields}}, \bibinfo {author}
  {\bibfnamefont {H.}~\bibnamefont {Sumiya}}, \bibinfo {author} {\bibfnamefont
  {S.}~\bibnamefont {Onoda}}, \bibinfo {author} {\bibfnamefont
  {J.}~\bibnamefont {Isoya}}, \bibinfo {author} {\bibfnamefont
  {E.}~\bibnamefont {Demler}}, \bibinfo {author} {\bibfnamefont
  {F.}~\bibnamefont {Jelezko}}, \bibinfo {author} {\bibfnamefont {N.~Y.}\
  \bibnamefont {Yao}},\ and\ \bibinfo {author} {\bibfnamefont {M.~D.}\
  \bibnamefont {Lukin}},\ }\href
  {https://doi.org/10.1103/PhysRevLett.118.093601} {\bibfield  {journal}
  {\bibinfo  {journal} {Phys. Rev. Lett.}\ }\textbf {\bibinfo {volume} {118}},\
  \bibinfo {pages} {093601} (\bibinfo {year} {2017}{\natexlab{a}})}\BibitemShut
  {NoStop}%
\bibitem [{\citenamefont {Zhou}\ \emph {et~al.}(2020)\citenamefont {Zhou},
  \citenamefont {Choi}, \citenamefont {Choi}, \citenamefont {Landig},
  \citenamefont {Douglas}, \citenamefont {Isoya}, \citenamefont {Jelezko},
  \citenamefont {Onoda}, \citenamefont {Sumiya}, \citenamefont {Cappellaro},
  \citenamefont {Knowles}, \citenamefont {Park},\ and\ \citenamefont
  {Lukin}}]{Zhou2020}%
  \BibitemOpen
  \bibfield  {author} {\bibinfo {author} {\bibfnamefont {H.}~\bibnamefont
  {Zhou}}, \bibinfo {author} {\bibfnamefont {J.}~\bibnamefont {Choi}}, \bibinfo
  {author} {\bibfnamefont {S.}~\bibnamefont {Choi}}, \bibinfo {author}
  {\bibfnamefont {R.}~\bibnamefont {Landig}}, \bibinfo {author} {\bibfnamefont
  {A.~M.}\ \bibnamefont {Douglas}}, \bibinfo {author} {\bibfnamefont
  {J.}~\bibnamefont {Isoya}}, \bibinfo {author} {\bibfnamefont
  {F.}~\bibnamefont {Jelezko}}, \bibinfo {author} {\bibfnamefont
  {S.}~\bibnamefont {Onoda}}, \bibinfo {author} {\bibfnamefont
  {H.}~\bibnamefont {Sumiya}}, \bibinfo {author} {\bibfnamefont
  {P.}~\bibnamefont {Cappellaro}}, \bibinfo {author} {\bibfnamefont {H.~S.}\
  \bibnamefont {Knowles}}, \bibinfo {author} {\bibfnamefont {H.}~\bibnamefont
  {Park}},\ and\ \bibinfo {author} {\bibfnamefont {M.~D.}\ \bibnamefont
  {Lukin}},\ }\href {https://doi.org/10.1103/PhysRevX.10.031003} {\bibfield
  {journal} {\bibinfo  {journal} {Phys. Rev. X}\ }\textbf {\bibinfo {volume}
  {10}},\ \bibinfo {pages} {031003} (\bibinfo {year} {2020})}\BibitemShut
  {NoStop}%
\bibitem [{\citenamefont {Davis}\ \emph {et~al.}(2023)\citenamefont {Davis},
  \citenamefont {Ye}, \citenamefont {Machado}, \citenamefont {Meynell},
  \citenamefont {Wu}, \citenamefont {Mittiga}, \citenamefont {Schenken},
  \citenamefont {Joos}, \citenamefont {Kobrin}, \citenamefont {Lyu},
  \citenamefont {Wang}, \citenamefont {Bluvstein}, \citenamefont {Choi},
  \citenamefont {Zu}, \citenamefont {Jayich},\ and\ \citenamefont
  {Yao}}]{Davis2023}%
  \BibitemOpen
  \bibfield  {author} {\bibinfo {author} {\bibfnamefont {E.~J.}\ \bibnamefont
  {Davis}}, \bibinfo {author} {\bibfnamefont {B.}~\bibnamefont {Ye}}, \bibinfo
  {author} {\bibfnamefont {F.}~\bibnamefont {Machado}}, \bibinfo {author}
  {\bibfnamefont {S.~A.}\ \bibnamefont {Meynell}}, \bibinfo {author}
  {\bibfnamefont {W.}~\bibnamefont {Wu}}, \bibinfo {author} {\bibfnamefont
  {T.}~\bibnamefont {Mittiga}}, \bibinfo {author} {\bibfnamefont
  {W.}~\bibnamefont {Schenken}}, \bibinfo {author} {\bibfnamefont
  {M.}~\bibnamefont {Joos}}, \bibinfo {author} {\bibfnamefont {B.}~\bibnamefont
  {Kobrin}}, \bibinfo {author} {\bibfnamefont {Y.}~\bibnamefont {Lyu}},
  \bibinfo {author} {\bibfnamefont {Z.}~\bibnamefont {Wang}}, \bibinfo {author}
  {\bibfnamefont {D.}~\bibnamefont {Bluvstein}}, \bibinfo {author}
  {\bibfnamefont {S.}~\bibnamefont {Choi}}, \bibinfo {author} {\bibfnamefont
  {C.}~\bibnamefont {Zu}}, \bibinfo {author} {\bibfnamefont {A.~C.~B.}\
  \bibnamefont {Jayich}},\ and\ \bibinfo {author} {\bibfnamefont {N.~Y.}\
  \bibnamefont {Yao}},\ }\href {https://doi.org/10.1038/s41567-023-01944-5}
  {\bibfield  {journal} {\bibinfo  {journal} {Nature Physics}\ }\textbf
  {\bibinfo {volume} {19}},\ \bibinfo {pages} {836} (\bibinfo {year}
  {2023})}\BibitemShut {NoStop}%
\bibitem [{\citenamefont {Hughes}\ \emph {et~al.}(2025)\citenamefont {Hughes},
  \citenamefont {Meynell}, \citenamefont {Wu}, \citenamefont {Parthasarathy},
  \citenamefont {Chen}, \citenamefont {Zhang}, \citenamefont {Wang},
  \citenamefont {Davis}, \citenamefont {Mukherjee}, \citenamefont {Yao},\ and\
  \citenamefont {Bleszynski~Jayich}}]{Hughes2025}%
  \BibitemOpen
  \bibfield  {author} {\bibinfo {author} {\bibfnamefont {L.~B.}\ \bibnamefont
  {Hughes}}, \bibinfo {author} {\bibfnamefont {S.~A.}\ \bibnamefont {Meynell}},
  \bibinfo {author} {\bibfnamefont {W.}~\bibnamefont {Wu}}, \bibinfo {author}
  {\bibfnamefont {S.}~\bibnamefont {Parthasarathy}}, \bibinfo {author}
  {\bibfnamefont {L.}~\bibnamefont {Chen}}, \bibinfo {author} {\bibfnamefont
  {Z.}~\bibnamefont {Zhang}}, \bibinfo {author} {\bibfnamefont
  {Z.}~\bibnamefont {Wang}}, \bibinfo {author} {\bibfnamefont {E.~J.}\
  \bibnamefont {Davis}}, \bibinfo {author} {\bibfnamefont {K.}~\bibnamefont
  {Mukherjee}}, \bibinfo {author} {\bibfnamefont {N.~Y.}\ \bibnamefont {Yao}},\
  and\ \bibinfo {author} {\bibfnamefont {A.~C.}\ \bibnamefont
  {Bleszynski~Jayich}},\ }\href {https://doi.org/10.1103/PhysRevX.15.021035}
  {\bibfield  {journal} {\bibinfo  {journal} {Phys. Rev. X}\ }\textbf {\bibinfo
  {volume} {15}},\ \bibinfo {pages} {021035} (\bibinfo {year}
  {2025})}\BibitemShut {NoStop}%
\bibitem [{\citenamefont {Kitagawa}\ and\ \citenamefont
  {Ueda}(1993)}]{Kitagawa1993}%
  \BibitemOpen
  \bibfield  {author} {\bibinfo {author} {\bibfnamefont {M.}~\bibnamefont
  {Kitagawa}}\ and\ \bibinfo {author} {\bibfnamefont {M.}~\bibnamefont
  {Ueda}},\ }\href {https://doi.org/10.1103/PhysRevA.47.5138} {\bibfield
  {journal} {\bibinfo  {journal} {Phys. Rev. A}\ }\textbf {\bibinfo {volume}
  {47}},\ \bibinfo {pages} {5138} (\bibinfo {year} {1993})}\BibitemShut
  {NoStop}%
\bibitem [{\citenamefont {Wineland}\ \emph {et~al.}(1992)\citenamefont
  {Wineland}, \citenamefont {Bollinger}, \citenamefont {Itano}, \citenamefont
  {Moore},\ and\ \citenamefont {Heinzen}}]{Wineland1992}%
  \BibitemOpen
  \bibfield  {author} {\bibinfo {author} {\bibfnamefont {D.~J.}\ \bibnamefont
  {Wineland}}, \bibinfo {author} {\bibfnamefont {J.~J.}\ \bibnamefont
  {Bollinger}}, \bibinfo {author} {\bibfnamefont {W.~M.}\ \bibnamefont
  {Itano}}, \bibinfo {author} {\bibfnamefont {F.~L.}\ \bibnamefont {Moore}},\
  and\ \bibinfo {author} {\bibfnamefont {D.~J.}\ \bibnamefont {Heinzen}},\
  }\href {https://doi.org/10.1103/PhysRevA.46.R6797} {\bibfield  {journal}
  {\bibinfo  {journal} {Phys. Rev. A}\ }\textbf {\bibinfo {volume} {46}},\
  \bibinfo {pages} {R6797} (\bibinfo {year} {1992})}\BibitemShut {NoStop}%
\bibitem [{\citenamefont {Meyer}\ \emph {et~al.}(2001)\citenamefont {Meyer},
  \citenamefont {Rowe}, \citenamefont {Kielpinski}, \citenamefont {Sackett},
  \citenamefont {Itano}, \citenamefont {Monroe},\ and\ \citenamefont
  {Wineland}}]{Meyer2001}%
  \BibitemOpen
  \bibfield  {author} {\bibinfo {author} {\bibfnamefont {V.}~\bibnamefont
  {Meyer}}, \bibinfo {author} {\bibfnamefont {M.~A.}\ \bibnamefont {Rowe}},
  \bibinfo {author} {\bibfnamefont {D.}~\bibnamefont {Kielpinski}}, \bibinfo
  {author} {\bibfnamefont {C.~A.}\ \bibnamefont {Sackett}}, \bibinfo {author}
  {\bibfnamefont {W.~M.}\ \bibnamefont {Itano}}, \bibinfo {author}
  {\bibfnamefont {C.}~\bibnamefont {Monroe}},\ and\ \bibinfo {author}
  {\bibfnamefont {D.~J.}\ \bibnamefont {Wineland}},\ }\href
  {https://doi.org/10.1103/PhysRevLett.86.5870} {\bibfield  {journal} {\bibinfo
   {journal} {Phys. Rev. Lett.}\ }\textbf {\bibinfo {volume} {86}},\ \bibinfo
  {pages} {5870} (\bibinfo {year} {2001})}\BibitemShut {NoStop}%
\bibitem [{\citenamefont {Gross}\ \emph {et~al.}(2010)\citenamefont {Gross},
  \citenamefont {Zibold}, \citenamefont {Nicklas}, \citenamefont {Est{\`e}ve},\
  and\ \citenamefont {Oberthaler}}]{Gross2010}%
  \BibitemOpen
  \bibfield  {author} {\bibinfo {author} {\bibfnamefont {C.}~\bibnamefont
  {Gross}}, \bibinfo {author} {\bibfnamefont {T.}~\bibnamefont {Zibold}},
  \bibinfo {author} {\bibfnamefont {E.}~\bibnamefont {Nicklas}}, \bibinfo
  {author} {\bibfnamefont {J.}~\bibnamefont {Est{\`e}ve}},\ and\ \bibinfo
  {author} {\bibfnamefont {M.~K.}\ \bibnamefont {Oberthaler}},\ }\href
  {https://doi.org/10.1038/nature08919} {\bibfield  {journal} {\bibinfo
  {journal} {Nature}\ }\textbf {\bibinfo {volume} {464}},\ \bibinfo {pages}
  {1165} (\bibinfo {year} {2010})}\BibitemShut {NoStop}%
\bibitem [{\citenamefont {Sewell}\ \emph {et~al.}(2012)\citenamefont {Sewell},
  \citenamefont {Koschorreck}, \citenamefont {Napolitano}, \citenamefont
  {Dubost}, \citenamefont {Behbood},\ and\ \citenamefont
  {Mitchell}}]{Sewell2012}%
  \BibitemOpen
  \bibfield  {author} {\bibinfo {author} {\bibfnamefont {R.~J.}\ \bibnamefont
  {Sewell}}, \bibinfo {author} {\bibfnamefont {M.}~\bibnamefont {Koschorreck}},
  \bibinfo {author} {\bibfnamefont {M.}~\bibnamefont {Napolitano}}, \bibinfo
  {author} {\bibfnamefont {B.}~\bibnamefont {Dubost}}, \bibinfo {author}
  {\bibfnamefont {N.}~\bibnamefont {Behbood}},\ and\ \bibinfo {author}
  {\bibfnamefont {M.~W.}\ \bibnamefont {Mitchell}},\ }\href
  {https://doi.org/10.1103/PhysRevLett.109.253605} {\bibfield  {journal}
  {\bibinfo  {journal} {Phys. Rev. Lett.}\ }\textbf {\bibinfo {volume} {109}},\
  \bibinfo {pages} {253605} (\bibinfo {year} {2012})}\BibitemShut {NoStop}%
\bibitem [{\citenamefont {Hosten}\ \emph
  {et~al.}(2016{\natexlab{a}})\citenamefont {Hosten}, \citenamefont {Engelsen},
  \citenamefont {Krishnakumar},\ and\ \citenamefont {Kasevich}}]{Hosten2016}%
  \BibitemOpen
  \bibfield  {author} {\bibinfo {author} {\bibfnamefont {O.}~\bibnamefont
  {Hosten}}, \bibinfo {author} {\bibfnamefont {N.~J.}\ \bibnamefont
  {Engelsen}}, \bibinfo {author} {\bibfnamefont {R.}~\bibnamefont
  {Krishnakumar}},\ and\ \bibinfo {author} {\bibfnamefont {M.~A.}\ \bibnamefont
  {Kasevich}},\ }\href {https://doi.org/10.1038/nature16176} {\bibfield
  {journal} {\bibinfo  {journal} {Nature}\ }\textbf {\bibinfo {volume} {529}},\
  \bibinfo {pages} {505} (\bibinfo {year} {2016}{\natexlab{a}})}\BibitemShut
  {NoStop}%
\bibitem [{\citenamefont {Block}\ \emph {et~al.}(2024)\citenamefont {Block},
  \citenamefont {Ye}, \citenamefont {Roberts}, \citenamefont {Chern},
  \citenamefont {Wu}, \citenamefont {Wang}, \citenamefont {Pollet},
  \citenamefont {Davis}, \citenamefont {Halperin},\ and\ \citenamefont
  {Yao}}]{Block2024}%
  \BibitemOpen
  \bibfield  {author} {\bibinfo {author} {\bibfnamefont {M.}~\bibnamefont
  {Block}}, \bibinfo {author} {\bibfnamefont {B.}~\bibnamefont {Ye}}, \bibinfo
  {author} {\bibfnamefont {B.}~\bibnamefont {Roberts}}, \bibinfo {author}
  {\bibfnamefont {S.}~\bibnamefont {Chern}}, \bibinfo {author} {\bibfnamefont
  {W.}~\bibnamefont {Wu}}, \bibinfo {author} {\bibfnamefont {Z.}~\bibnamefont
  {Wang}}, \bibinfo {author} {\bibfnamefont {L.}~\bibnamefont {Pollet}},
  \bibinfo {author} {\bibfnamefont {E.~J.}\ \bibnamefont {Davis}}, \bibinfo
  {author} {\bibfnamefont {B.~I.}\ \bibnamefont {Halperin}},\ and\ \bibinfo
  {author} {\bibfnamefont {N.~Y.}\ \bibnamefont {Yao}},\ }\href
  {https://doi.org/10.1038/s41567-024-02562-5} {\bibfield  {journal} {\bibinfo
  {journal} {Nat. Phys.}\ }\textbf {\bibinfo {volume} {20}},\ \bibinfo {pages}
  {1575} (\bibinfo {year} {2024})}\BibitemShut {NoStop}%
\bibitem [{\citenamefont {Wu}\ \emph {et~al.}(2025)\citenamefont {Wu},
  \citenamefont {Davis}, \citenamefont {Hughes}, \citenamefont {Ye},
  \citenamefont {Wang}, \citenamefont {Kufel}, \citenamefont {Ono},
  \citenamefont {Meynell}, \citenamefont {Block}, \citenamefont {Liu},
  \citenamefont {Yang}, \citenamefont {Bleszynski~Jayich},\ and\ \citenamefont
  {Yao}}]{Wu2025}%
  \BibitemOpen
  \bibfield  {author} {\bibinfo {author} {\bibfnamefont {W.}~\bibnamefont
  {Wu}}, \bibinfo {author} {\bibfnamefont {E.~J.}\ \bibnamefont {Davis}},
  \bibinfo {author} {\bibfnamefont {L.~B.}\ \bibnamefont {Hughes}}, \bibinfo
  {author} {\bibfnamefont {B.}~\bibnamefont {Ye}}, \bibinfo {author}
  {\bibfnamefont {Z.}~\bibnamefont {Wang}}, \bibinfo {author} {\bibfnamefont
  {D.}~\bibnamefont {Kufel}}, \bibinfo {author} {\bibfnamefont
  {T.}~\bibnamefont {Ono}}, \bibinfo {author} {\bibfnamefont {S.~A.}\
  \bibnamefont {Meynell}}, \bibinfo {author} {\bibfnamefont {M.}~\bibnamefont
  {Block}}, \bibinfo {author} {\bibfnamefont {C.}~\bibnamefont {Liu}}, \bibinfo
  {author} {\bibfnamefont {H.}~\bibnamefont {Yang}}, \bibinfo {author}
  {\bibfnamefont {A.~C.}\ \bibnamefont {Bleszynski~Jayich}},\ and\ \bibinfo
  {author} {\bibfnamefont {N.~Y.}\ \bibnamefont {Yao}},\ }\href
  {https://doi.org/10.1038/s41586-025-09524-8} {\bibfield  {journal} {\bibinfo
  {journal} {Nature}\ }\textbf {\bibinfo {volume} {646}},\ \bibinfo {pages}
  {74} (\bibinfo {year} {2025})}\BibitemShut {NoStop}%
\bibitem [{\citenamefont {Kaplan-Lipkin}\ \emph {et~al.}(2025)\citenamefont
  {Kaplan-Lipkin}, \citenamefont {Crowley}, \citenamefont {Hallén},
  \citenamefont {Wang}, \citenamefont {Wu}, \citenamefont {Chern},
  \citenamefont {Laumann}, \citenamefont {Pollet},\ and\ \citenamefont
  {Yao}}]{KaplanLipkin2025}%
  \BibitemOpen
  \bibfield  {author} {\bibinfo {author} {\bibfnamefont {A.}~\bibnamefont
  {Kaplan-Lipkin}}, \bibinfo {author} {\bibfnamefont {P.~J.~D.}\ \bibnamefont
  {Crowley}}, \bibinfo {author} {\bibfnamefont {J.~N.}\ \bibnamefont
  {Hallén}}, \bibinfo {author} {\bibfnamefont {Z.}~\bibnamefont {Wang}},
  \bibinfo {author} {\bibfnamefont {W.}~\bibnamefont {Wu}}, \bibinfo {author}
  {\bibfnamefont {S.}~\bibnamefont {Chern}}, \bibinfo {author} {\bibfnamefont
  {C.~R.}\ \bibnamefont {Laumann}}, \bibinfo {author} {\bibfnamefont
  {L.}~\bibnamefont {Pollet}},\ and\ \bibinfo {author} {\bibfnamefont {N.~Y.}\
  \bibnamefont {Yao}},\ }\href {https://doi.org/10.48550/arXiv.2512.19781}
  {\bibinfo {title} {Theory of {Scalable} {Spin} {Squeezing} with {Disordered}
  {Quantum} {Dipoles}}} (\bibinfo {year} {2025}),\ \bibinfo {note}
  {arXiv:2512.19781 [quant-ph]}\BibitemShut {NoStop}%
\bibitem [{SM()}]{SM}%
  \BibitemOpen
  \href@noop {} {}\bibinfo {note} {See Supplemental Material.}\BibitemShut
  {Stop}%
\bibitem [{\citenamefont {Gruber}\ \emph {et~al.}(1997)\citenamefont {Gruber},
  \citenamefont {Dräbenstedt}, \citenamefont {Tietz}, \citenamefont {Fleury},
  \citenamefont {Wrachtrup},\ and\ \citenamefont {von
  Borczyskowski}}]{Gruber1997}%
  \BibitemOpen
  \bibfield  {author} {\bibinfo {author} {\bibfnamefont {A.}~\bibnamefont
  {Gruber}}, \bibinfo {author} {\bibfnamefont {A.}~\bibnamefont
  {Dräbenstedt}}, \bibinfo {author} {\bibfnamefont {C.}~\bibnamefont {Tietz}},
  \bibinfo {author} {\bibfnamefont {L.}~\bibnamefont {Fleury}}, \bibinfo
  {author} {\bibfnamefont {J.}~\bibnamefont {Wrachtrup}},\ and\ \bibinfo
  {author} {\bibfnamefont {C.}~\bibnamefont {von Borczyskowski}},\ }\href
  {https://doi.org/10.1126/science.276.5321.2012} {\bibfield  {journal}
  {\bibinfo  {journal} {Science}\ }\textbf {\bibinfo {volume} {276}},\ \bibinfo
  {pages} {2012} (\bibinfo {year} {1997})}\BibitemShut {NoStop}%
\bibitem [{\citenamefont {Manson}\ \emph {et~al.}(2006)\citenamefont {Manson},
  \citenamefont {Harrison},\ and\ \citenamefont {Sellars}}]{Manson2006}%
  \BibitemOpen
  \bibfield  {author} {\bibinfo {author} {\bibfnamefont {N.~B.}\ \bibnamefont
  {Manson}}, \bibinfo {author} {\bibfnamefont {J.~P.}\ \bibnamefont
  {Harrison}},\ and\ \bibinfo {author} {\bibfnamefont {M.~J.}\ \bibnamefont
  {Sellars}},\ }\href {https://doi.org/10.1103/PhysRevB.74.104303} {\bibfield
  {journal} {\bibinfo  {journal} {Phys. Rev. B}\ }\textbf {\bibinfo {volume}
  {74}},\ \bibinfo {pages} {104303} (\bibinfo {year} {2006})}\BibitemShut
  {NoStop}%
\bibitem [{Note1()}]{Note1}%
  \BibitemOpen
  \bibinfo {note} {The moderate field requirement here circumvents the
  substantial infrastructure demands of tesla-scale superconducting magnets,
  keeping the protocol compatible with the use of ensemble NVs as a practical
  sensing platform.}\BibitemShut {Stop}%
\bibitem [{Note2()}]{Note2}%
  \BibitemOpen
  \bibinfo {note} {The variation in the number of available readouts $k$ among
  spots is mainly due to the optimal laser power, which itself depends on
  spot-specific parameters such the density (see~\cite {SM}). In particular,
  spot A exhibited a significantly faster contrast decay than spots B and C,
  despite their similar NV densities. We attribute this to the electronic-spin
  initialization time, normalized by the optimal readout time, being
  approximately twice as long in spot A as in spots B and C, resulting in
  poorer electronic-spin polarization after each readout and, consequently,
  faster loss of the nuclear-spin memory.}\BibitemShut {Stop}%
\bibitem [{Note3()}]{Note3}%
  \BibitemOpen
  \bibinfo {note} {The photon count signal $n = \DOTSB \sum@ \slimits@ _i w_i
  n_i$, where $w_i$ accounts for contrast decay arising from nuclear-spin
  depolarization and $n_i$ is the photon count from the $i$th repetitive
  readout.}\BibitemShut {Stop}%
\bibitem [{Note4()}]{Note4}%
  \BibitemOpen
  \bibinfo {note} {We note that this cannot arise from the photon shot noise
  subtraction since $\sigma _{\protect \rm PSN}^2$ exhibits only one
  oscillation per Rabi cycle.}\BibitemShut {Stop}%
\bibitem [{\citenamefont {Schloss}\ \emph {et~al.}(2018)\citenamefont
  {Schloss}, \citenamefont {Barry}, \citenamefont {Turner},\ and\ \citenamefont
  {Walsworth}}]{Schloss2018}%
  \BibitemOpen
  \bibfield  {author} {\bibinfo {author} {\bibfnamefont {J.~M.}\ \bibnamefont
  {Schloss}}, \bibinfo {author} {\bibfnamefont {J.~F.}\ \bibnamefont {Barry}},
  \bibinfo {author} {\bibfnamefont {M.~J.}\ \bibnamefont {Turner}},\ and\
  \bibinfo {author} {\bibfnamefont {R.~L.}\ \bibnamefont {Walsworth}},\ }\href
  {https://doi.org/10.1103/PhysRevApplied.10.034044} {\bibfield  {journal}
  {\bibinfo  {journal} {Phys. Rev. Appl.}\ }\textbf {\bibinfo {volume} {10}},\
  \bibinfo {pages} {034044} (\bibinfo {year} {2018})}\BibitemShut {NoStop}%
\bibitem [{\citenamefont {Barry}\ \emph {et~al.}(2016)\citenamefont {Barry},
  \citenamefont {Turner}, \citenamefont {Schloss}, \citenamefont {Glenn},
  \citenamefont {Song}, \citenamefont {Lukin}, \citenamefont {Park},\ and\
  \citenamefont {Walsworth}}]{Barry2016}%
  \BibitemOpen
  \bibfield  {author} {\bibinfo {author} {\bibfnamefont {J.~F.}\ \bibnamefont
  {Barry}}, \bibinfo {author} {\bibfnamefont {M.~J.}\ \bibnamefont {Turner}},
  \bibinfo {author} {\bibfnamefont {J.~M.}\ \bibnamefont {Schloss}}, \bibinfo
  {author} {\bibfnamefont {D.~R.}\ \bibnamefont {Glenn}}, \bibinfo {author}
  {\bibfnamefont {Y.}~\bibnamefont {Song}}, \bibinfo {author} {\bibfnamefont
  {M.~D.}\ \bibnamefont {Lukin}}, \bibinfo {author} {\bibfnamefont
  {H.}~\bibnamefont {Park}},\ and\ \bibinfo {author} {\bibfnamefont {R.~L.}\
  \bibnamefont {Walsworth}},\ }\href {https://doi.org/10.1073/pnas.1601513113}
  {\bibfield  {journal} {\bibinfo  {journal} {Proceedings of the National
  Academy of Sciences}\ }\textbf {\bibinfo {volume} {113}},\ \bibinfo {pages}
  {14133} (\bibinfo {year} {2016})}\BibitemShut {NoStop}%
\bibitem [{\citenamefont {Chatzidrosos}\ \emph {et~al.}(2017)\citenamefont
  {Chatzidrosos}, \citenamefont {Wickenbrock}, \citenamefont {Bougas},
  \citenamefont {Leefer}, \citenamefont {Wu}, \citenamefont {Jensen},
  \citenamefont {Dumeige},\ and\ \citenamefont {Budker}}]{Chatzidrosos2017}%
  \BibitemOpen
  \bibfield  {author} {\bibinfo {author} {\bibfnamefont {G.}~\bibnamefont
  {Chatzidrosos}}, \bibinfo {author} {\bibfnamefont {A.}~\bibnamefont
  {Wickenbrock}}, \bibinfo {author} {\bibfnamefont {L.}~\bibnamefont {Bougas}},
  \bibinfo {author} {\bibfnamefont {N.}~\bibnamefont {Leefer}}, \bibinfo
  {author} {\bibfnamefont {T.}~\bibnamefont {Wu}}, \bibinfo {author}
  {\bibfnamefont {K.}~\bibnamefont {Jensen}}, \bibinfo {author} {\bibfnamefont
  {Y.}~\bibnamefont {Dumeige}},\ and\ \bibinfo {author} {\bibfnamefont
  {D.}~\bibnamefont {Budker}},\ }\href
  {https://doi.org/10.1103/PhysRevApplied.8.044019} {\bibfield  {journal}
  {\bibinfo  {journal} {Phys. Rev. Appl.}\ }\textbf {\bibinfo {volume} {8}},\
  \bibinfo {pages} {044019} (\bibinfo {year} {2017})}\BibitemShut {NoStop}%
\bibitem [{\citenamefont {Le~Sage}\ \emph {et~al.}(2012)\citenamefont
  {Le~Sage}, \citenamefont {Pham}, \citenamefont {Bar-Gill}, \citenamefont
  {Belthangady}, \citenamefont {Lukin}, \citenamefont {Yacoby},\ and\
  \citenamefont {Walsworth}}]{LeSage2012}%
  \BibitemOpen
  \bibfield  {author} {\bibinfo {author} {\bibfnamefont {D.}~\bibnamefont
  {Le~Sage}}, \bibinfo {author} {\bibfnamefont {L.~M.}\ \bibnamefont {Pham}},
  \bibinfo {author} {\bibfnamefont {N.}~\bibnamefont {Bar-Gill}}, \bibinfo
  {author} {\bibfnamefont {C.}~\bibnamefont {Belthangady}}, \bibinfo {author}
  {\bibfnamefont {M.~D.}\ \bibnamefont {Lukin}}, \bibinfo {author}
  {\bibfnamefont {A.}~\bibnamefont {Yacoby}},\ and\ \bibinfo {author}
  {\bibfnamefont {R.~L.}\ \bibnamefont {Walsworth}},\ }\href
  {https://doi.org/10.1103/PhysRevB.85.121202} {\bibfield  {journal} {\bibinfo
  {journal} {Phys. Rev. B}\ }\textbf {\bibinfo {volume} {85}},\ \bibinfo
  {pages} {121202(R)} (\bibinfo {year} {2012})}\BibitemShut {NoStop}%
\bibitem [{\citenamefont {Gullion}\ \emph {et~al.}(1990)\citenamefont
  {Gullion}, \citenamefont {Baker},\ and\ \citenamefont
  {Conradi}}]{Gullion1990}%
  \BibitemOpen
  \bibfield  {author} {\bibinfo {author} {\bibfnamefont {T.}~\bibnamefont
  {Gullion}}, \bibinfo {author} {\bibfnamefont {D.~B.}\ \bibnamefont {Baker}},\
  and\ \bibinfo {author} {\bibfnamefont {M.~S.}\ \bibnamefont {Conradi}},\
  }\href {https://doi.org/10.1016/0022-2364(90)90331-3} {\bibfield  {journal}
  {\bibinfo  {journal} {J. Magn. Reson.}\ }\textbf {\bibinfo {volume} {89}},\
  \bibinfo {pages} {479} (\bibinfo {year} {1990})}\BibitemShut {NoStop}%
\bibitem [{\citenamefont {de~Lange}\ \emph {et~al.}(2010)\citenamefont
  {de~Lange}, \citenamefont {Wang}, \citenamefont {Rist{\`e}}, \citenamefont
  {Dobrovitski},\ and\ \citenamefont {Hanson}}]{deLange2010}%
  \BibitemOpen
  \bibfield  {author} {\bibinfo {author} {\bibfnamefont {G.}~\bibnamefont
  {de~Lange}}, \bibinfo {author} {\bibfnamefont {Z.~H.}\ \bibnamefont {Wang}},
  \bibinfo {author} {\bibfnamefont {D.}~\bibnamefont {Rist{\`e}}}, \bibinfo
  {author} {\bibfnamefont {V.~V.}\ \bibnamefont {Dobrovitski}},\ and\ \bibinfo
  {author} {\bibfnamefont {R.}~\bibnamefont {Hanson}},\ }\href
  {https://doi.org/10.1126/science.1192739} {\bibfield  {journal} {\bibinfo
  {journal} {Science}\ }\textbf {\bibinfo {volume} {330}},\ \bibinfo {pages}
  {60} (\bibinfo {year} {2010})}\BibitemShut {NoStop}%
\bibitem [{\citenamefont {Choi}\ \emph {et~al.}(2020)\citenamefont {Choi},
  \citenamefont {Zhou}, \citenamefont {Knowles}, \citenamefont {Landig},
  \citenamefont {Choi},\ and\ \citenamefont {Lukin}}]{Choi2020}%
  \BibitemOpen
  \bibfield  {author} {\bibinfo {author} {\bibfnamefont {J.}~\bibnamefont
  {Choi}}, \bibinfo {author} {\bibfnamefont {H.}~\bibnamefont {Zhou}}, \bibinfo
  {author} {\bibfnamefont {H.~S.}\ \bibnamefont {Knowles}}, \bibinfo {author}
  {\bibfnamefont {R.}~\bibnamefont {Landig}}, \bibinfo {author} {\bibfnamefont
  {S.}~\bibnamefont {Choi}},\ and\ \bibinfo {author} {\bibfnamefont {M.~D.}\
  \bibnamefont {Lukin}},\ }\href {https://doi.org/10.1103/PhysRevX.10.031002}
  {\bibfield  {journal} {\bibinfo  {journal} {Phys. Rev. X}\ }\textbf {\bibinfo
  {volume} {10}},\ \bibinfo {pages} {031002} (\bibinfo {year}
  {2020})}\BibitemShut {NoStop}%
\bibitem [{\citenamefont {Schweigler}\ \emph {et~al.}(2017)\citenamefont
  {Schweigler}, \citenamefont {Kasper}, \citenamefont {Erne}, \citenamefont
  {Mazets}, \citenamefont {Rauer}, \citenamefont {Cataldini}, \citenamefont
  {Langen}, \citenamefont {Gasenzer}, \citenamefont {Berges},\ and\
  \citenamefont {Schmiedmayer}}]{Schweigler2017}%
  \BibitemOpen
  \bibfield  {author} {\bibinfo {author} {\bibfnamefont {T.}~\bibnamefont
  {Schweigler}}, \bibinfo {author} {\bibfnamefont {V.}~\bibnamefont {Kasper}},
  \bibinfo {author} {\bibfnamefont {S.}~\bibnamefont {Erne}}, \bibinfo {author}
  {\bibfnamefont {I.}~\bibnamefont {Mazets}}, \bibinfo {author} {\bibfnamefont
  {B.}~\bibnamefont {Rauer}}, \bibinfo {author} {\bibfnamefont
  {F.}~\bibnamefont {Cataldini}}, \bibinfo {author} {\bibfnamefont
  {T.}~\bibnamefont {Langen}}, \bibinfo {author} {\bibfnamefont
  {T.}~\bibnamefont {Gasenzer}}, \bibinfo {author} {\bibfnamefont
  {J.}~\bibnamefont {Berges}},\ and\ \bibinfo {author} {\bibfnamefont
  {J.}~\bibnamefont {Schmiedmayer}},\ }\href
  {https://doi.org/10.1038/nature22310} {\bibfield  {journal} {\bibinfo
  {journal} {Nature}\ }\textbf {\bibinfo {volume} {545}},\ \bibinfo {pages}
  {323} (\bibinfo {year} {2017})}\BibitemShut {NoStop}%
\bibitem [{\citenamefont {Pr{\"u}fer}\ \emph {et~al.}(2020)\citenamefont
  {Pr{\"u}fer}, \citenamefont {Zache}, \citenamefont {Kunkel}, \citenamefont
  {Lannig}, \citenamefont {Bonnin}, \citenamefont {Strobel}, \citenamefont
  {Berges},\ and\ \citenamefont {Oberthaler}}]{Pruefer2020}%
  \BibitemOpen
  \bibfield  {author} {\bibinfo {author} {\bibfnamefont {M.}~\bibnamefont
  {Pr{\"u}fer}}, \bibinfo {author} {\bibfnamefont {T.~V.}\ \bibnamefont
  {Zache}}, \bibinfo {author} {\bibfnamefont {P.}~\bibnamefont {Kunkel}},
  \bibinfo {author} {\bibfnamefont {S.}~\bibnamefont {Lannig}}, \bibinfo
  {author} {\bibfnamefont {A.}~\bibnamefont {Bonnin}}, \bibinfo {author}
  {\bibfnamefont {H.}~\bibnamefont {Strobel}}, \bibinfo {author} {\bibfnamefont
  {J.}~\bibnamefont {Berges}},\ and\ \bibinfo {author} {\bibfnamefont {M.~K.}\
  \bibnamefont {Oberthaler}},\ }\href
  {https://doi.org/10.1038/s41567-020-0933-6} {\bibfield  {journal} {\bibinfo
  {journal} {Nat. Phys.}\ }\textbf {\bibinfo {volume} {16}},\ \bibinfo {pages}
  {1012} (\bibinfo {year} {2020})}\BibitemShut {NoStop}%
\bibitem [{\citenamefont {Schweigler}\ \emph {et~al.}(2021)\citenamefont
  {Schweigler}, \citenamefont {Gluza}, \citenamefont {Tajik}, \citenamefont
  {Sotiriadis}, \citenamefont {Cataldini}, \citenamefont {Ji}, \citenamefont
  {M{\o}ller}, \citenamefont {Sabino}, \citenamefont {Rauer}, \citenamefont
  {Eisert},\ and\ \citenamefont {Schmiedmayer}}]{Schweigler2021}%
  \BibitemOpen
  \bibfield  {author} {\bibinfo {author} {\bibfnamefont {T.}~\bibnamefont
  {Schweigler}}, \bibinfo {author} {\bibfnamefont {M.}~\bibnamefont {Gluza}},
  \bibinfo {author} {\bibfnamefont {M.}~\bibnamefont {Tajik}}, \bibinfo
  {author} {\bibfnamefont {S.}~\bibnamefont {Sotiriadis}}, \bibinfo {author}
  {\bibfnamefont {F.}~\bibnamefont {Cataldini}}, \bibinfo {author}
  {\bibfnamefont {S.-C.}\ \bibnamefont {Ji}}, \bibinfo {author} {\bibfnamefont
  {F.~S.}\ \bibnamefont {M{\o}ller}}, \bibinfo {author} {\bibfnamefont
  {J.}~\bibnamefont {Sabino}}, \bibinfo {author} {\bibfnamefont
  {B.}~\bibnamefont {Rauer}}, \bibinfo {author} {\bibfnamefont
  {J.}~\bibnamefont {Eisert}},\ and\ \bibinfo {author} {\bibfnamefont
  {J.}~\bibnamefont {Schmiedmayer}},\ }\href
  {https://doi.org/10.1038/s41567-020-01139-2} {\bibfield  {journal} {\bibinfo
  {journal} {Nat. Phys.}\ }\textbf {\bibinfo {volume} {17}},\ \bibinfo {pages}
  {559} (\bibinfo {year} {2021})}\BibitemShut {NoStop}%
\bibitem [{\citenamefont {Strobel}\ \emph {et~al.}(2014)\citenamefont
  {Strobel}, \citenamefont {Muessel}, \citenamefont {Linnemann}, \citenamefont
  {Zibold}, \citenamefont {Hume}, \citenamefont {Pezz{\`e}}, \citenamefont
  {Smerzi},\ and\ \citenamefont {Oberthaler}}]{Strobel2014}%
  \BibitemOpen
  \bibfield  {author} {\bibinfo {author} {\bibfnamefont {H.}~\bibnamefont
  {Strobel}}, \bibinfo {author} {\bibfnamefont {W.}~\bibnamefont {Muessel}},
  \bibinfo {author} {\bibfnamefont {D.}~\bibnamefont {Linnemann}}, \bibinfo
  {author} {\bibfnamefont {T.}~\bibnamefont {Zibold}}, \bibinfo {author}
  {\bibfnamefont {D.~B.}\ \bibnamefont {Hume}}, \bibinfo {author}
  {\bibfnamefont {L.}~\bibnamefont {Pezz{\`e}}}, \bibinfo {author}
  {\bibfnamefont {A.}~\bibnamefont {Smerzi}},\ and\ \bibinfo {author}
  {\bibfnamefont {M.~K.}\ \bibnamefont {Oberthaler}},\ }\href
  {https://doi.org/10.1126/science.1250147} {\bibfield  {journal} {\bibinfo
  {journal} {Science}\ }\textbf {\bibinfo {volume} {345}},\ \bibinfo {pages}
  {424} (\bibinfo {year} {2014})}\BibitemShut {NoStop}%
\bibitem [{\citenamefont {Hosten}\ \emph
  {et~al.}(2016{\natexlab{b}})\citenamefont {Hosten}, \citenamefont
  {Krishnakumar}, \citenamefont {Engelsen},\ and\ \citenamefont
  {Kasevich}}]{HostenMagnification2016}%
  \BibitemOpen
  \bibfield  {author} {\bibinfo {author} {\bibfnamefont {O.}~\bibnamefont
  {Hosten}}, \bibinfo {author} {\bibfnamefont {R.}~\bibnamefont
  {Krishnakumar}}, \bibinfo {author} {\bibfnamefont {N.~J.}\ \bibnamefont
  {Engelsen}},\ and\ \bibinfo {author} {\bibfnamefont {M.~A.}\ \bibnamefont
  {Kasevich}},\ }\href {https://doi.org/10.1126/science.aaf3397} {\bibfield
  {journal} {\bibinfo  {journal} {Science}\ }\textbf {\bibinfo {volume}
  {352}},\ \bibinfo {pages} {1552} (\bibinfo {year}
  {2016}{\natexlab{b}})}\BibitemShut {NoStop}%
\bibitem [{\citenamefont {Choi}\ \emph
  {et~al.}(2017{\natexlab{b}})\citenamefont {Choi}, \citenamefont {Choi},
  \citenamefont {Landig}, \citenamefont {Kucsko}, \citenamefont {Zhou},
  \citenamefont {Isoya}, \citenamefont {Jelezko}, \citenamefont {Onoda},
  \citenamefont {Sumiya}, \citenamefont {Khemani}, \citenamefont {von
  Keyserlingk}, \citenamefont {Yao}, \citenamefont {Demler},\ and\
  \citenamefont {Lukin}}]{ChoiDTC2017}%
  \BibitemOpen
  \bibfield  {author} {\bibinfo {author} {\bibfnamefont {S.}~\bibnamefont
  {Choi}}, \bibinfo {author} {\bibfnamefont {J.}~\bibnamefont {Choi}}, \bibinfo
  {author} {\bibfnamefont {R.}~\bibnamefont {Landig}}, \bibinfo {author}
  {\bibfnamefont {G.}~\bibnamefont {Kucsko}}, \bibinfo {author} {\bibfnamefont
  {H.}~\bibnamefont {Zhou}}, \bibinfo {author} {\bibfnamefont {J.}~\bibnamefont
  {Isoya}}, \bibinfo {author} {\bibfnamefont {F.}~\bibnamefont {Jelezko}},
  \bibinfo {author} {\bibfnamefont {S.}~\bibnamefont {Onoda}}, \bibinfo
  {author} {\bibfnamefont {H.}~\bibnamefont {Sumiya}}, \bibinfo {author}
  {\bibfnamefont {V.}~\bibnamefont {Khemani}}, \bibinfo {author} {\bibfnamefont
  {C.}~\bibnamefont {von Keyserlingk}}, \bibinfo {author} {\bibfnamefont
  {N.~Y.}\ \bibnamefont {Yao}}, \bibinfo {author} {\bibfnamefont
  {E.}~\bibnamefont {Demler}},\ and\ \bibinfo {author} {\bibfnamefont {M.~D.}\
  \bibnamefont {Lukin}},\ }\href {https://doi.org/10.1038/nature21426}
  {\bibfield  {journal} {\bibinfo  {journal} {Nature}\ }\textbf {\bibinfo
  {volume} {543}},\ \bibinfo {pages} {221} (\bibinfo {year}
  {2017}{\natexlab{b}})}\BibitemShut {NoStop}%
\bibitem [{\citenamefont {Kucsko}\ \emph {et~al.}(2018)\citenamefont {Kucsko},
  \citenamefont {Choi}, \citenamefont {Choi}, \citenamefont {Maurer},
  \citenamefont {Zhou}, \citenamefont {Landig}, \citenamefont {Sumiya},
  \citenamefont {Onoda}, \citenamefont {Isoya}, \citenamefont {Jelezko},
  \citenamefont {Demler}, \citenamefont {Yao},\ and\ \citenamefont
  {Lukin}}]{Kucsko2018}%
  \BibitemOpen
  \bibfield  {author} {\bibinfo {author} {\bibfnamefont {G.}~\bibnamefont
  {Kucsko}}, \bibinfo {author} {\bibfnamefont {S.}~\bibnamefont {Choi}},
  \bibinfo {author} {\bibfnamefont {J.}~\bibnamefont {Choi}}, \bibinfo {author}
  {\bibfnamefont {P.~C.}\ \bibnamefont {Maurer}}, \bibinfo {author}
  {\bibfnamefont {H.}~\bibnamefont {Zhou}}, \bibinfo {author} {\bibfnamefont
  {R.}~\bibnamefont {Landig}}, \bibinfo {author} {\bibfnamefont
  {H.}~\bibnamefont {Sumiya}}, \bibinfo {author} {\bibfnamefont
  {S.}~\bibnamefont {Onoda}}, \bibinfo {author} {\bibfnamefont
  {J.}~\bibnamefont {Isoya}}, \bibinfo {author} {\bibfnamefont
  {F.}~\bibnamefont {Jelezko}}, \bibinfo {author} {\bibfnamefont
  {E.}~\bibnamefont {Demler}}, \bibinfo {author} {\bibfnamefont {N.~Y.}\
  \bibnamefont {Yao}},\ and\ \bibinfo {author} {\bibfnamefont {M.~D.}\
  \bibnamefont {Lukin}},\ }\href
  {https://doi.org/10.1103/PhysRevLett.121.023601} {\bibfield  {journal}
  {\bibinfo  {journal} {Phys. Rev. Lett.}\ }\textbf {\bibinfo {volume} {121}},\
  \bibinfo {pages} {023601} (\bibinfo {year} {2018})}\BibitemShut {NoStop}%
\bibitem [{\citenamefont {Zu}\ \emph {et~al.}(2021)\citenamefont {Zu},
  \citenamefont {Machado}, \citenamefont {Ye}, \citenamefont {Choi},
  \citenamefont {Kobrin}, \citenamefont {Mittiga}, \citenamefont {Hsieh},
  \citenamefont {Bhattacharyya}, \citenamefont {Markham}, \citenamefont
  {Twitchen}, \citenamefont {Jarmola}, \citenamefont {Budker}, \citenamefont
  {Laumann}, \citenamefont {Moore},\ and\ \citenamefont {Yao}}]{Zu2021}%
  \BibitemOpen
  \bibfield  {author} {\bibinfo {author} {\bibfnamefont {C.}~\bibnamefont
  {Zu}}, \bibinfo {author} {\bibfnamefont {F.}~\bibnamefont {Machado}},
  \bibinfo {author} {\bibfnamefont {B.}~\bibnamefont {Ye}}, \bibinfo {author}
  {\bibfnamefont {S.}~\bibnamefont {Choi}}, \bibinfo {author} {\bibfnamefont
  {B.}~\bibnamefont {Kobrin}}, \bibinfo {author} {\bibfnamefont
  {T.}~\bibnamefont {Mittiga}}, \bibinfo {author} {\bibfnamefont
  {S.}~\bibnamefont {Hsieh}}, \bibinfo {author} {\bibfnamefont
  {P.}~\bibnamefont {Bhattacharyya}}, \bibinfo {author} {\bibfnamefont
  {M.}~\bibnamefont {Markham}}, \bibinfo {author} {\bibfnamefont
  {D.}~\bibnamefont {Twitchen}}, \bibinfo {author} {\bibfnamefont
  {A.}~\bibnamefont {Jarmola}}, \bibinfo {author} {\bibfnamefont
  {D.}~\bibnamefont {Budker}}, \bibinfo {author} {\bibfnamefont {C.~R.}\
  \bibnamefont {Laumann}}, \bibinfo {author} {\bibfnamefont {J.~E.}\
  \bibnamefont {Moore}},\ and\ \bibinfo {author} {\bibfnamefont {N.~Y.}\
  \bibnamefont {Yao}},\ }\href {https://doi.org/10.1038/s41586-021-03763-1}
  {\bibfield  {journal} {\bibinfo  {journal} {Nature}\ }\textbf {\bibinfo
  {volume} {597}},\ \bibinfo {pages} {45} (\bibinfo {year} {2021})}\BibitemShut
  {NoStop}%
\bibitem [{\citenamefont {Rovny}\ \emph {et~al.}(2024)\citenamefont {Rovny},
  \citenamefont {Gopalakrishnan}, \citenamefont {Bleszynski~Jayich},
  \citenamefont {Maletinsky}, \citenamefont {Demler},\ and\ \citenamefont
  {de~Leon}}]{Rovny2024}%
  \BibitemOpen
  \bibfield  {author} {\bibinfo {author} {\bibfnamefont {J.}~\bibnamefont
  {Rovny}}, \bibinfo {author} {\bibfnamefont {S.}~\bibnamefont
  {Gopalakrishnan}}, \bibinfo {author} {\bibfnamefont {A.~C.}\ \bibnamefont
  {Bleszynski~Jayich}}, \bibinfo {author} {\bibfnamefont {P.}~\bibnamefont
  {Maletinsky}}, \bibinfo {author} {\bibfnamefont {E.}~\bibnamefont {Demler}},\
  and\ \bibinfo {author} {\bibfnamefont {N.~P.}\ \bibnamefont {de~Leon}},\
  }\href {https://doi.org/10.1038/s42254-024-00775-4} {\bibfield  {journal}
  {\bibinfo  {journal} {Nat. Rev. Phys.}\ }\textbf {\bibinfo {volume} {6}},\
  \bibinfo {pages} {753} (\bibinfo {year} {2024})}\BibitemShut {NoStop}%
\end{thebibliography}%


\begin{thebibliography}{8}%
\makeatletter
\providecommand \@ifxundefined [1]{%
 \@ifx{#1\undefined}
}%
\providecommand \@ifnum [1]{%
 \ifnum #1\expandafter \@firstoftwo
 \else \expandafter \@secondoftwo
 \fi
}%
\providecommand \@ifx [1]{%
 \ifx #1\expandafter \@firstoftwo
 \else \expandafter \@secondoftwo
 \fi
}%
\providecommand \natexlab [1]{#1}%
\providecommand \enquote  [1]{``#1''}%
\providecommand \bibnamefont  [1]{#1}%
\providecommand \bibfnamefont [1]{#1}%
\providecommand \citenamefont [1]{#1}%
\providecommand \href@noop [0]{\@secondoftwo}%
\providecommand \href [0]{\begingroup \@sanitize@url \@href}%
\providecommand \@href[1]{\@@startlink{#1}\@@href}%
\providecommand \@@href[1]{\endgroup#1\@@endlink}%
\providecommand \@sanitize@url [0]{\catcode `\\12\catcode `\$12\catcode
  `\&12\catcode `\#12\catcode `\^12\catcode `\_12\catcode `\%12\relax}%
\providecommand \@@startlink[1]{}%
\providecommand \@@endlink[0]{}%
\providecommand \url  [0]{\begingroup\@sanitize@url \@url }%
\providecommand \@url [1]{\endgroup\@href {#1}{\urlprefix }}%
\providecommand \urlprefix  [0]{URL }%
\providecommand \Eprint [0]{\href }%
\providecommand \doibase [0]{https://doi.org/}%
\providecommand \selectlanguage [0]{\@gobble}%
\providecommand \bibinfo  [0]{\@secondoftwo}%
\providecommand \bibfield  [0]{\@secondoftwo}%
\providecommand \translation [1]{[#1]}%
\providecommand \BibitemOpen [0]{}%
\providecommand \bibitemStop [0]{}%
\providecommand \bibitemNoStop [0]{.\EOS\space}%
\providecommand \EOS [0]{\spacefactor3000\relax}%
\providecommand \BibitemShut  [1]{\csname bibitem#1\endcsname}%
\let\auto@bib@innerbib\@empty
\bibitem [{\citenamefont {Hughes}\ \emph {et~al.}(2025)\citenamefont {Hughes},
  \citenamefont {Meynell}, \citenamefont {Wu}, \citenamefont {Parthasarathy},
  \citenamefont {Chen}, \citenamefont {Zhang}, \citenamefont {Wang},
  \citenamefont {Davis}, \citenamefont {Mukherjee}, \citenamefont {Yao},\ and\
  \citenamefont {Bleszynski~Jayich}}]{SMref:Hughes2025}%
  \BibitemOpen
  \bibfield  {author} {\bibinfo {author} {\bibfnamefont {L.~B.}\ \bibnamefont
  {Hughes}}, \bibinfo {author} {\bibfnamefont {S.~A.}\ \bibnamefont {Meynell}},
  \bibinfo {author} {\bibfnamefont {W.}~\bibnamefont {Wu}}, \bibinfo {author}
  {\bibfnamefont {S.}~\bibnamefont {Parthasarathy}}, \bibinfo {author}
  {\bibfnamefont {L.}~\bibnamefont {Chen}}, \bibinfo {author} {\bibfnamefont
  {Z.}~\bibnamefont {Zhang}}, \bibinfo {author} {\bibfnamefont
  {Z.}~\bibnamefont {Wang}}, \bibinfo {author} {\bibfnamefont {E.~J.}\
  \bibnamefont {Davis}}, \bibinfo {author} {\bibfnamefont {K.}~\bibnamefont
  {Mukherjee}}, \bibinfo {author} {\bibfnamefont {N.~Y.}\ \bibnamefont {Yao}},\
  and\ \bibinfo {author} {\bibfnamefont {A.~C.}\ \bibnamefont
  {Bleszynski~Jayich}},\ }\href {https://doi.org/10.1103/PhysRevX.15.021035}
  {\bibfield  {journal} {\bibinfo  {journal} {Phys. Rev. X}\ }\textbf {\bibinfo
  {volume} {15}},\ \bibinfo {pages} {021035} (\bibinfo {year}
  {2025})}\BibitemShut {NoStop}%
\bibitem [{\citenamefont {Starkey}(2019)}]{SMref:Labscript}%
  \BibitemOpen
  \bibfield  {author} {\bibinfo {author} {\bibfnamefont {P.~T.}\ \bibnamefont
  {Starkey}},\ }\emph {\bibinfo {title} {A software framework for control and
  automation of precisely timed experiments}},\ \href
  {https://doi.org/10.26180/8637200.v1} {Ph.D. thesis},\ \bibinfo  {school}
  {Monash University} (\bibinfo {year} {2019})\BibitemShut {NoStop}%
\bibitem [{\citenamefont {Tetienne}\ \emph {et~al.}(2012)\citenamefont
  {Tetienne}, \citenamefont {Rondin}, \citenamefont {Spinicelli}, \citenamefont
  {Chipaux}, \citenamefont {Debuisschert}, \citenamefont {Roch},\ and\
  \citenamefont {Jacques}}]{SMref:Tetienne2012}%
  \BibitemOpen
  \bibfield  {author} {\bibinfo {author} {\bibfnamefont {J.-P.}\ \bibnamefont
  {Tetienne}}, \bibinfo {author} {\bibfnamefont {L.}~\bibnamefont {Rondin}},
  \bibinfo {author} {\bibfnamefont {P.}~\bibnamefont {Spinicelli}}, \bibinfo
  {author} {\bibfnamefont {M.}~\bibnamefont {Chipaux}}, \bibinfo {author}
  {\bibfnamefont {T.}~\bibnamefont {Debuisschert}}, \bibinfo {author}
  {\bibfnamefont {J.-F.}\ \bibnamefont {Roch}},\ and\ \bibinfo {author}
  {\bibfnamefont {V.}~\bibnamefont {Jacques}},\ }\href
  {https://doi.org/10.1088/1367-2630/14/10/103033} {\bibfield  {journal}
  {\bibinfo  {journal} {New Journal of Physics}\ }\textbf {\bibinfo {volume}
  {14}},\ \bibinfo {pages} {103033} (\bibinfo {year} {2012})}\BibitemShut
  {NoStop}%
\bibitem [{\citenamefont {Jiang}\ \emph {et~al.}(2009)\citenamefont {Jiang},
  \citenamefont {Hodges}, \citenamefont {Maze}, \citenamefont {Maurer},
  \citenamefont {Taylor}, \citenamefont {Cory}, \citenamefont {Hemmer},
  \citenamefont {Walsworth}, \citenamefont {Yacoby}, \citenamefont {Zibrov},\
  and\ \citenamefont {Lukin}}]{SMref:Jiang2009}%
  \BibitemOpen
  \bibfield  {author} {\bibinfo {author} {\bibfnamefont {L.}~\bibnamefont
  {Jiang}}, \bibinfo {author} {\bibfnamefont {J.~S.}\ \bibnamefont {Hodges}},
  \bibinfo {author} {\bibfnamefont {J.~R.}\ \bibnamefont {Maze}}, \bibinfo
  {author} {\bibfnamefont {P.}~\bibnamefont {Maurer}}, \bibinfo {author}
  {\bibfnamefont {J.~M.}\ \bibnamefont {Taylor}}, \bibinfo {author}
  {\bibfnamefont {D.~G.}\ \bibnamefont {Cory}}, \bibinfo {author}
  {\bibfnamefont {P.~R.}\ \bibnamefont {Hemmer}}, \bibinfo {author}
  {\bibfnamefont {R.~L.}\ \bibnamefont {Walsworth}}, \bibinfo {author}
  {\bibfnamefont {A.}~\bibnamefont {Yacoby}}, \bibinfo {author} {\bibfnamefont
  {A.~S.}\ \bibnamefont {Zibrov}},\ and\ \bibinfo {author} {\bibfnamefont
  {M.~D.}\ \bibnamefont {Lukin}},\ }\href
  {https://doi.org/10.1126/science.1176496} {\bibfield  {journal} {\bibinfo
  {journal} {Science}\ }\textbf {\bibinfo {volume} {326}},\ \bibinfo {pages}
  {267} (\bibinfo {year} {2009})}\BibitemShut {NoStop}%
\bibitem [{\citenamefont {Nakamura}\ \emph {et~al.}(2022)\citenamefont
  {Nakamura}, \citenamefont {Watanabe}, \citenamefont {Sumiya}, \citenamefont
  {Itoh}, \citenamefont {Sasaki}, \citenamefont {Ishi-Hayase},\ and\
  \citenamefont {Kobayashi}}]{SMref:Nakamura2022}%
  \BibitemOpen
  \bibfield  {author} {\bibinfo {author} {\bibfnamefont {Y.}~\bibnamefont
  {Nakamura}}, \bibinfo {author} {\bibfnamefont {H.}~\bibnamefont {Watanabe}},
  \bibinfo {author} {\bibfnamefont {H.}~\bibnamefont {Sumiya}}, \bibinfo
  {author} {\bibfnamefont {K.~M.}\ \bibnamefont {Itoh}}, \bibinfo {author}
  {\bibfnamefont {K.}~\bibnamefont {Sasaki}}, \bibinfo {author} {\bibfnamefont
  {J.}~\bibnamefont {Ishi-Hayase}},\ and\ \bibinfo {author} {\bibfnamefont
  {K.}~\bibnamefont {Kobayashi}},\ }\href {https://doi.org/10.1063/5.0090450}
  {\bibfield  {journal} {\bibinfo  {journal} {AIP Advances}\ }\textbf {\bibinfo
  {volume} {12}},\ \bibinfo {pages} {055215} (\bibinfo {year}
  {2022})}\BibitemShut {NoStop}%
\bibitem [{\citenamefont {Hopper}\ \emph {et~al.}(2018)\citenamefont {Hopper},
  \citenamefont {Shulevitz},\ and\ \citenamefont {Bassett}}]{SMref:Hopper2018}%
  \BibitemOpen
  \bibfield  {author} {\bibinfo {author} {\bibfnamefont {D.~A.}\ \bibnamefont
  {Hopper}}, \bibinfo {author} {\bibfnamefont {H.~J.}\ \bibnamefont
  {Shulevitz}},\ and\ \bibinfo {author} {\bibfnamefont {L.~C.}\ \bibnamefont
  {Bassett}},\ }\bibfield  {journal} {\bibinfo  {journal} {Micromachines}\
  }\textbf {\bibinfo {volume} {9}},\ \href {https://doi.org/10.3390/mi9090437}
  {10.3390/mi9090437} (\bibinfo {year} {2018})\BibitemShut {NoStop}%
\bibitem [{\citenamefont {Wu}\ \emph {et~al.}(2025)\citenamefont {Wu},
  \citenamefont {Davis}, \citenamefont {Hughes}, \citenamefont {Ye},
  \citenamefont {Wang}, \citenamefont {Kufel}, \citenamefont {Ono},
  \citenamefont {Meynell}, \citenamefont {Block}, \citenamefont {Liu},
  \citenamefont {Yang}, \citenamefont {Bleszynski~Jayich},\ and\ \citenamefont
  {Yao}}]{SMref:Wu2025}%
  \BibitemOpen
  \bibfield  {author} {\bibinfo {author} {\bibfnamefont {W.}~\bibnamefont
  {Wu}}, \bibinfo {author} {\bibfnamefont {E.~J.}\ \bibnamefont {Davis}},
  \bibinfo {author} {\bibfnamefont {L.~B.}\ \bibnamefont {Hughes}}, \bibinfo
  {author} {\bibfnamefont {B.}~\bibnamefont {Ye}}, \bibinfo {author}
  {\bibfnamefont {Z.}~\bibnamefont {Wang}}, \bibinfo {author} {\bibfnamefont
  {D.}~\bibnamefont {Kufel}}, \bibinfo {author} {\bibfnamefont
  {T.}~\bibnamefont {Ono}}, \bibinfo {author} {\bibfnamefont {S.~A.}\
  \bibnamefont {Meynell}}, \bibinfo {author} {\bibfnamefont {M.}~\bibnamefont
  {Block}}, \bibinfo {author} {\bibfnamefont {C.}~\bibnamefont {Liu}}, \bibinfo
  {author} {\bibfnamefont {H.}~\bibnamefont {Yang}}, \bibinfo {author}
  {\bibfnamefont {A.~C.}\ \bibnamefont {Bleszynski~Jayich}},\ and\ \bibinfo
  {author} {\bibfnamefont {N.~Y.}\ \bibnamefont {Yao}},\ }\href
  {https://doi.org/10.1038/s41586-025-09524-8} {\bibfield  {journal} {\bibinfo
  {journal} {Nature}\ }\textbf {\bibinfo {volume} {646}},\ \bibinfo {pages}
  {74} (\bibinfo {year} {2025})}\BibitemShut {NoStop}%
\bibitem [{\citenamefont {Davis}\ \emph {et~al.}(2023)\citenamefont {Davis},
  \citenamefont {Ye}, \citenamefont {Machado}, \citenamefont {Meynell},
  \citenamefont {Wu}, \citenamefont {Mittiga}, \citenamefont {Schenken},
  \citenamefont {Joos}, \citenamefont {Kobrin}, \citenamefont {Lyu},
  \citenamefont {Wang}, \citenamefont {Bluvstein}, \citenamefont {Choi},
  \citenamefont {Zu}, \citenamefont {Jayich},\ and\ \citenamefont
  {Yao}}]{SMref:Davis2023}%
  \BibitemOpen
  \bibfield  {author} {\bibinfo {author} {\bibfnamefont {E.~J.}\ \bibnamefont
  {Davis}}, \bibinfo {author} {\bibfnamefont {B.}~\bibnamefont {Ye}}, \bibinfo
  {author} {\bibfnamefont {F.}~\bibnamefont {Machado}}, \bibinfo {author}
  {\bibfnamefont {S.~A.}\ \bibnamefont {Meynell}}, \bibinfo {author}
  {\bibfnamefont {W.}~\bibnamefont {Wu}}, \bibinfo {author} {\bibfnamefont
  {T.}~\bibnamefont {Mittiga}}, \bibinfo {author} {\bibfnamefont
  {W.}~\bibnamefont {Schenken}}, \bibinfo {author} {\bibfnamefont
  {M.}~\bibnamefont {Joos}}, \bibinfo {author} {\bibfnamefont {B.}~\bibnamefont
  {Kobrin}}, \bibinfo {author} {\bibfnamefont {Y.}~\bibnamefont {Lyu}},
  \bibinfo {author} {\bibfnamefont {Z.}~\bibnamefont {Wang}}, \bibinfo {author}
  {\bibfnamefont {D.}~\bibnamefont {Bluvstein}}, \bibinfo {author}
  {\bibfnamefont {S.}~\bibnamefont {Choi}}, \bibinfo {author} {\bibfnamefont
  {C.}~\bibnamefont {Zu}}, \bibinfo {author} {\bibfnamefont {A.~C.~B.}\
  \bibnamefont {Jayich}},\ and\ \bibinfo {author} {\bibfnamefont {N.~Y.}\
  \bibnamefont {Yao}},\ }\href {https://doi.org/10.1038/s41567-023-01944-5}
  {\bibfield  {journal} {\bibinfo  {journal} {Nature Physics}\ }\textbf
  {\bibinfo {volume} {19}},\ \bibinfo {pages} {836} (\bibinfo {year}
  {2023})}\BibitemShut {NoStop}%
\end{thebibliography}%

\newcommand{\supplementincluded}{}
\let\supplementfinish\relax
\let\suppcite\citesupp
\newcommand{\supplementbibliography}{%
  \bibliographystylesupp{apsrev4-2}%
  \bibliographysupp{suppmat}%
}

\clearpage
\onecolumngrid
\begingroup

\makeatletter
\newcommand{\supplementtableofcontents}{%
  \par\bigskip
  {\centering\large\bfseries Contents\par}%
  \medskip
  \@starttoc{stoc}%
}
\let\supplement@addcontentsline\addcontentsline
\def\supplement@tocname{toc}
\renewcommand{\addcontentsline}[3]{%
  \def\supplement@requestedfile{#1}%
  \ifx\supplement@requestedfile\supplement@tocname
    \supplement@addcontentsline{stoc}{#2}{#3}%
  \else
    \supplement@addcontentsline{#1}{#2}{#3}%
  \fi
}
\makeatother

\setcounter{page}{1}
\setcounter{section}{0}
\setcounter{subsection}{0}
\setcounter{figure}{0}
\setcounter{table}{0}
\setcounter{equation}{0}
\renewcommand{\thesection}{S\arabic{section}}
\renewcommand{\thesubsection}{\thesection.\arabic{subsection}}
\renewcommand{\thefigure}{S\arabic{figure}}
\renewcommand{\thetable}{S\arabic{table}}
\renewcommand{\theequation}{S\arabic{equation}}
\renewcommand{\theHsection}{supplementary.section.\arabic{section}}
\renewcommand{\theHsubsection}{supplementary.subsection.\arabic{section}.\arabic{subsection}}
\renewcommand{\theHfigure}{supplementary.figure.\arabic{figure}}
\renewcommand{\theHtable}{supplementary.table.\arabic{table}}
\renewcommand{\theHequation}{supplementary.equation.\arabic{equation}}

\begin{center}
  {\large\bfseries Supplemental Material for\\[0.4em]
  Direct Observation of Dipolar-Driven Anisotropic Quantum Projection Noise in a Solid-State Spin Ensemble}
\end{center}

\vspace{0.5em}
\supplementtableofcontents
\vspace{0.5em}

\section{Experimental Setup}

\subsection{Sample properties}

The NV centers reside in a (111)-oriented, isotopically purified $^{12}$C diamond epilayer grown by plasma-enhanced chemical vapor deposition. During growth, $^{15}$N gas is briefly introduced to form a nitrogen delta-doped layer, confining the NVs to a sheet of $\lesssim 7$~nm thickness, estimated by secondary-ion mass spectrometry. Subsequent electron-beam irradiation and annealing create 18 circular regions of high NV density within this layer, with diameters of $2$--$12~\mu\mathrm{m}$. Ref.~\suppcite{SMref:Hughes2025} describes the sample growth, irradiation, and characterization in detail. We use four of these regions, labeled A, B, C and D. Confocal images of the regions are shown in Fig.~\ref{fig:supp-scan}.

\begin{figure}[b]
\centering
\includegraphics[width=0.3\textwidth]{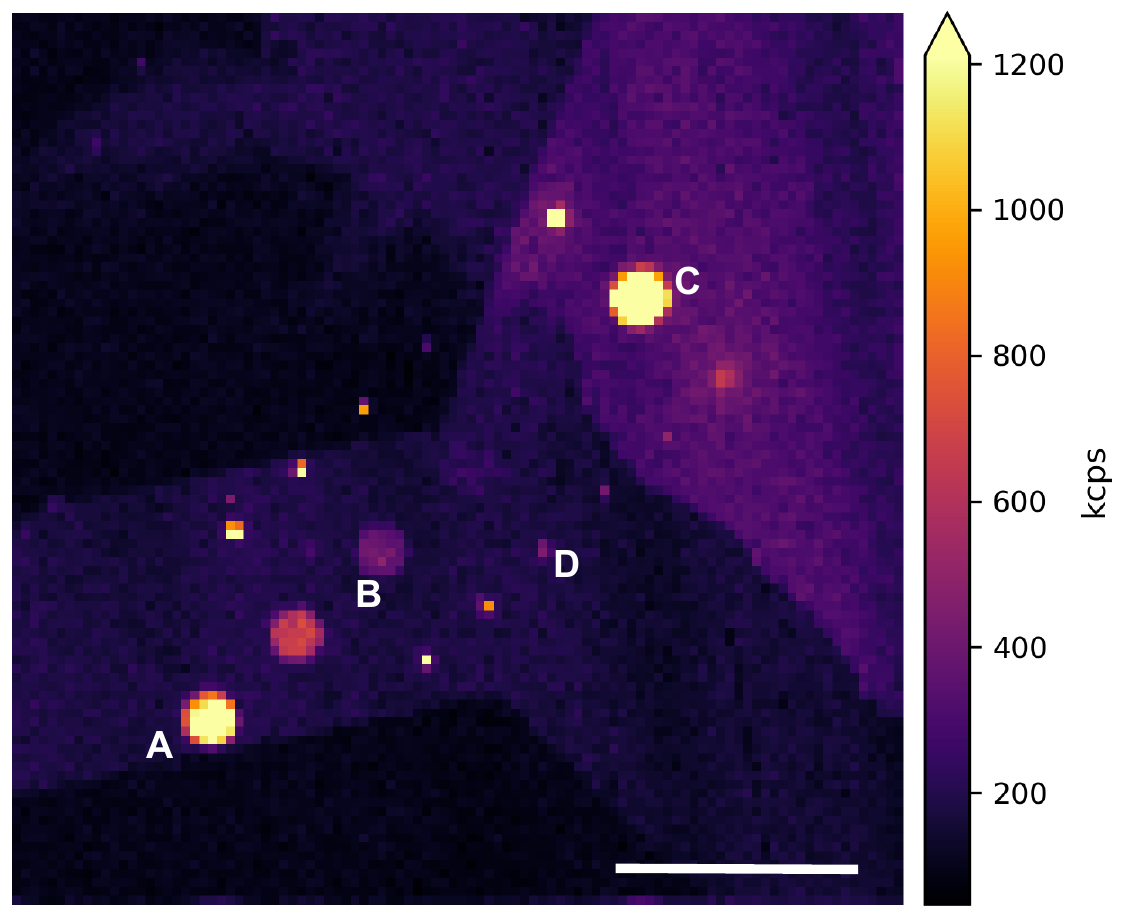}
\caption{The four circular regions of high NV density created by electron-beam irradiation, labeled here and in the main text as A, B, C, and D are visible in the confocal scan. The scale bar is $50~\mu\mathrm{m}$.
}
\label{fig:supp-scan}
\end{figure}

\subsection{Optical setup, microwave control, and data acquisition}

The sample sits in a home-built $4f$ scanning confocal microscope at room temperature. A $532~\mathrm{nm}$ laser, focused through an air objective (Nikon Plan Apo 60X, NA 0.95), optically initializes the NV electronic spins and enables fluorescence readout. For conventional fluorescence readout, the emitted photons from the NVs are separated from the green excitation light by a dichroic filter and then collected by fiber-coupled single-photon counters (Excelitas SPCM-AQRH-12-FC).

A Xilinx ZCU111 FPGA board generates both the microwave signal that drives the NV electronic-spin transition and the radio-frequency signal that drives the $^{15}$N nuclear-spin transition. The two signals are amplified separately (Mini-Circuits ZHL-50W-63+ and LZY-22+, respectively), combined by a diplexer (WDK8+7-DC-2690/3500-10000-60S3), and delivered to the sample. DC blocks and band-pass filters are added to suppress unwanted noise.

Experimental control and data acquisition are performed with the labscript suite, a Python-based framework for hardware-timed experiments~\suppcite{SMref:Labscript}.

\subsection{Magnetic-field generation and alignment}
An \ce{Nd2Fe14B} permanent ring magnet generates the magnetic field. The magnet has an inner diameter of $45~\si{mm}$, an outer diameter of $107~\si{mm}$, and a thickness of $17~\si{mm}$.  
We align the field normal to the sample plane, along the [111] crystallographic direction, solely by moving the magnet relative to the fixed sample. Accurate alignment is important because the nuclear-spin relaxation timescale $T_{1n}$ (see the main text) depends on the field orientation. At the high fields used here, the NV photoluminescence (PL) depends strongly on the angle between the applied field and the NV symmetry axis, providing an in~situ alignment signal~\suppcite{SMref:Tetienne2012}. We scan the magnet position with a three-axis motorized translation stage (Thorlabs PT3-Z9) while monitoring the NV PL; the PL maximum then defines the best-aligned configuration.

\subsection{Temperature control}

Temperature changes alter the permanent magnet's field and can therefore degrade the C$_n$NOT$_e$ gate. These temperature-induced gate-fidelity fluctuations add readout noise, increasing $\sigma_R$.
To estimate the required temperature stability, note that the CNOT gate uses a weak microwave signal that drives only one of the two hyperfine groups separated by $A_{\parallel} = 3.03~\si{MHz}$. The reversible temperature coefficient of remanence for NdFeB magnets is $-0.12~\%/\si{K}$, corresponding at $\sim 2700~\si{G}$ (the magnetic field at the position of the sample) to a field drift of $\sim -3.2~\si{G}/\si{K}$ and an NV transition-frequency shift of $\sim 9.1~\si{MHz}/\si{K}$. A fluctuation of $\sim 340~\si{mK}$ is thus sufficient to shift the hyperfine-selective transition frequency to that of the unwanted hyperfine group; the magnet temperature must remain stable to well below this scale throughout the measurements.

We therefore actively stabilize the temperature around the setup with a closed-loop proportional--integral--derivative (PID) controller based on an OMRON E5CC-QX2ABM-000. A volume of $\sim 1~\si{m}^3$ around the magnet and the sample is isolated from the outside by acrylic boards with a small hole for fast thermalization with the outside space. The PID controller takes its input from a Pt100 resistance thermometer placed near the sample and the magnet, and switches a solid-state relay (Crydom DC60D20) that controls the current through resistive-film heaters (Turboflex 45-555678) bonded to a metallic plate. The controller can automatically tune its gains and hold the temperature within $\pm 0.01~\si{K}$ of the setpoint. The corresponding residual field drift is $\lesssim 0.03~\si{G}$, and the NV transition-frequency drift is $\lesssim 0.1~\si{MHz}$. In addition, we periodically monitor the NV transition frequency and adjust the frequency of the generated MW signals during the experiment.

\section{Repetitive Readout}

\subsection{Post-processing of the repetitive readout data}
\label{sec:postprocess}

Each cycle of the repetitive readout produces a train of photon counts $n_1, n_2, \ldots, n_{2\ell}$, one for each of the $2\ell$ readout shots, which post-processing combines into a single signal $n$ by taking a weighted sum:
\begin{equation}\label{eq:weighted-sum}
  n = \sum_{i=1}^{2\ell} w_i\, n_i .
\end{equation}
Here we derive the optimal weights and the resulting $\sigma_R$, first for a general weighted sum and then for the differential protocol used in the experiment, following Refs.~\suppcite{SMref:Jiang2009, SMref:Nakamura2022}.

Consider one cycle of the repetitive readout of a single NV. The electronic spin is prepared in $\cos(\theta/2)\ket{0}_e + \sin(\theta/2)\ket{-1}_e$ and swapped onto the nuclear memory, and the nuclear-spin state is then read out over the sequence of single shots indexed by $i$. Let $a_i$ and $b_i$ be the mean number of photons detected in the $i$-th shot when the prepared electronic spin state is in $\ket{0} (\theta = 0)$ and $\ket{-1} (\theta = \pi)$, respectively. We obtain these reference traces separately by initializing the spin into each state. Note that the magnitude of the contrast $c_i \equiv a_i - b_i$ decays with the shot index $i$, because the nuclear memory depolarizes over its lifetime $T_{1n}$. 

Every shot in a cycle reads out the same projected spin state, so the photon counts across shots are correlated. With probability $\cos^2(\theta/2)$ the projection returns $\ket{0}$, and each shot $i$ then contributes an independent Poisson count of mean $a_i$; with probability $\sin^2(\theta/2)$ the projection returns $\ket{-1}$, with mean $b_i$.
The signal $n$ therefore has mean
\begin{equation}
  \expval{n} = \cos^2\!\left(\tfrac{\theta}{2}\right)\sum_i w_i a_i + \sin^2\!\left(\tfrac{\theta}{2}\right)\sum_i w_i b_i
\end{equation}
and variance
\begin{equation}
  \sigma_n^2 = \sum_i w_i^2\left[\cos^2\!\left(\tfrac{\theta}{2}\right)a_i + \sin^2\!\left(\tfrac{\theta}{2}\right)b_i\right]
  + \cos^2\!\left(\tfrac{\theta}{2}\right)\sin^2\!\left(\tfrac{\theta}{2}\right)\left(\sum_i w_i c_i\right)^2 .
\end{equation}
The first term is the photon shot noise, where each Poisson count has a variance equal to its mean, and the second term is the quantum projection noise of the spin. Here, we assume that the total noise consists only of these two contributions, and we neglect any additional classical noise.

The readout noise $\sigma_R$ is defined by~\suppcite{SMref:Hopper2018},
\begin{equation}
  \sigma_R = \left. \frac{\sigma_n}{\left|\partial\expval{n}/\partial\theta\right|} \right|_{\theta = \pi/2},
\end{equation}
which is equal to unity when the total noise reaches the quantum projection noise limit (no photon shot noise). Evaluating the two expressions above at $\theta = \pi/2$ gives %
\begin{equation}\label{eq:sigmaR-weights}
  \sigma_R = \sqrt{1 + \frac{2\sum_i w_i^2 (a_i + b_i)}{\left(\sum_i w_i c_i\right)^2}}.
\end{equation}

One can tune $w_i$ such that the readout is optimal, i.e., $\sigma_R$ is minimized~\suppcite{SMref:Jiang2009}. By the Cauchy--Schwarz inequality,
\begin{equation}
  \left(\sum_i w_i c_i\right)^2 \le \left(\sum_i w_i^2 (a_i + b_i)\right)\left(\sum_i \frac{c_i^2}{a_i + b_i}\right),
\end{equation}
with equality when
\begin{equation}\label{eq:optimal-weights}
  w_i \propto \frac{c_i}{a_i + b_i} = \frac{a_i - b_i}{a_i + b_i} .
\end{equation}
The optimal weight of each shot is therefore set by its contrast; shots taken late in the cycle, where the memory has partly depolarized, are down-weighted. Inserting the optimal weights into Eq.~\eqref{eq:sigmaR-weights} gives the smallest achievable readout noise
\begin{equation}\label{eq:sigmaR-min-sum}
  \sigma_R^{\min} = \sqrt{1 + \frac{2}{\sum_i c_i^2/(a_i + b_i)}} .
\end{equation}
In practice, we measure the reference traces $a_i$ and $b_i$ in separate calibration runs with electronic spins initialized to $\ket{0}_e$ and $\ket{-1}_e$, respectively, and evaluate the weights. We note that any additional classical noise can drive the weighting derived above away from the optimum.

In our implementation of repetitive readout, we pair each shot with the subsequent shot addressed by the complementary CNOT gate to form the differential signal
\begin{equation}\label{eq:diff-signal}
  n = \sum_{i=1}^{\ell} w_i \left( n_{2i-1} - n_{2i} \right),
\end{equation}
which is Eq.~\eqref{eq:weighted-sum} restricted to weights of equal magnitude and opposite sign within each pair, $w_{2i-1} = -w_{2i}$. The benefit of enforcing the pairing is robustness. Each pair difference $n_{2i-1} - n_{2i}$ cancels a photon-count background noise common to the two shots, so that, for example, slow drifts of the laser power or the temperature are rejected pair by pair.
Substituting the constraint $w_{2i-1} = -w_{2i}$ into Eq.~\eqref{eq:sigmaR-weights} gives
\begin{equation}\label{eq:sigmaR-diff}
  \sigma_R = \sqrt{1 + \frac{2\sum_i w_i^2\, \sigma_i}{\left(\sum_i w_i \delta_i\right)^2}} ,
\end{equation}
where $\delta_i \equiv c_{2i-1} - c_{2i},\sigma_i \equiv a_{2i-1} + b_{2i-1} + a_{2i} + b_{2i}$ are the contrast and the mean total photon number of the $i$-th pair. Note that the single-shot contrasts within a pair carry opposite signs.
The same Cauchy--Schwarz argument therefore yields the optimal pair weights $w_i \propto \frac{\delta_i}{\sigma_i}$ and the minimal readout noise of the differential scheme,
\begin{equation}\label{eq:sigmaR-min-diff}
  \sigma_R^{\min} = \sqrt{1 + \frac{2}{\sum_i \delta_i^2/\sigma_i}} .
\end{equation}
The pairing constraint can only increase $\sigma_R$ relative to the unconstrained optimum of Eq.~\eqref{eq:sigmaR-min-sum}. 
In the experiment, because $T_{1n}$ spans many shots, the memory decays only marginally between the two shots of a pair, so the difference in $\sigma_{R}$ is negligible. For spot C, we further tuned the weighting so that the shot-to-shot correlation in a repetitive readout sequence (which is considered to be due to laser-induced charge effects) is suppressed.

\subsection{Classical simulation of the repetitive readout process}
\label{sec:classical-sim}

We describe our method for extracting the system size from the measured residual variance $\sigma_{\rm res}^2$ of the differential repetitive readout. Let us start from the simplest case, where the residual variance of the $N$-spin product state $[\cos(\theta/2)\ket{0}_e+\sin(\theta/2)\ket{-1}_e]^{\otimes N}$ is given only by the quantum projection noise, i.e., $\sigma_{\rm res}^2 = \sigma_{\rm QPN}^2 = N\sin^2\theta/4$. In this case, the system size $N$ can be extracted directly from the amplitude of the oscillation of $\sigma_{\rm res}^2$ as a function of $\theta$. In practice, however, the measured residual variance differs from the quantum projection noise; in particular, the finite hyperfine linewidth, the incomplete electronic repolarization between shots, and the depolarization of the nuclear memory over the repeated readouts can contribute additional noise. Thus, we simulate the full differential repetitive-readout sequence classically, using the independently measured properties of each spot as input and keeping the system size $N$ as the fit parameter.

The simulation propagates uncorrelated spins through the experimental pulse sequence as Monte Carlo trajectories. Each spin begins in a thermal electronic state and carries its own hyperfine detuning, drawn from a Lorentzian whose half-width is the measured hyperfine linewidth. Microwave and radio-frequency pulses act on the detuned spins with the exact detuning-dependent Rabi transfer probability, which determines the CNOT gate infidelity. Each green pulse repolarizes both the electronic and nuclear spins with probability $1 - e^{-T_{R, \rm single}/\tau}$, where $T_{R, \rm single}$ is the duration of the laser pulse applied in a single conventional readout, and $\tau$ is the characteristic time scale of the repolarization process, obtained from a separate measurement; for the electronic spin, $\tau$ corresponds to the $T_1$ time under laser illumination, and for the nuclear spin, $\tau$ corresponds to $T_{1n}$.

At the beginning of each laser pulse, the ensemble emits a Poisson-distributed number of photons. The photon counts are combined with the same weights $w_i$ that are applied to the data [Eq.~\eqref{eq:diff-signal}]. The residual noise $\sigma_{\rm res}^2$ is then calculated in the same manner as in the experiment. We scan the system size and, at each value of $N$, quantify the agreement between simulation and experiment through the loss
\begin{equation}\label{eq:sim-loss}
  \mathcal{L}(N) = \sum_{y} \frac{\left\langle \left(y_{\rm sim} - y_{\rm exp}\right)^2 \right\rangle_\theta}{\left\langle y_{\rm exp} \right\rangle_\theta^2}
  + \left( \frac{A_{\rm sim} - A_{\rm exp}}{A_{\rm exp}} \right)^2,
\end{equation}
where $y \in \{\sigma_{\rm tot}^2, \sigma_{\rm PSN}^2, \sigma_{\rm res}^2\}$ runs over the total variance, the photon shot noise, and the residual variance, $\langle \cdot \rangle_\theta$ denotes the average over the measured preparation angles, and $A$ is the oscillation amplitude of $\sigma_{\rm res}^2$, defined as its standard deviation across the angles.

At the best-fit $N$, the simulation reproduces the full $\theta$ dependence of $\sigma_{\rm res}^2$ to $10$--$20\%$ rms, and the total variance and the photon shot noise to $\lesssim 2\%$; the simulated $\sigma_{\rm res}^2$ curves are the dashed lines shown in Fig.~\ref{fig:fig-2}(b) of the main text. The loss develops a sharp interior minimum as a function of $N$ (Fig.~\ref{fig:supp_readoutsim}), which gives the optimal system size for each spot,
\begin{equation}
  N^{\mathrm{A}} = 41(11), \qquad N^{\mathrm{B}} = 48(6), \qquad N^{\mathrm{C}} = 69(8).
\end{equation}
The quoted uncertainty is set by the quality of the fit rather than by statistics. The best fit reproduces the measured curves to $10$--$20\%$ rms, and values of $N$ near the optimum cannot be distinguished until they degrade this agreement well beyond that level. We therefore accept every $N$ for which the rms deviation stays below twice its optimal value (shaded ranges in Fig.~\ref{fig:supp_readoutsim}), and quote the resulting range as the uncertainty.

\begin{figure}[t]
\centering
\includegraphics[width=0.4\linewidth]{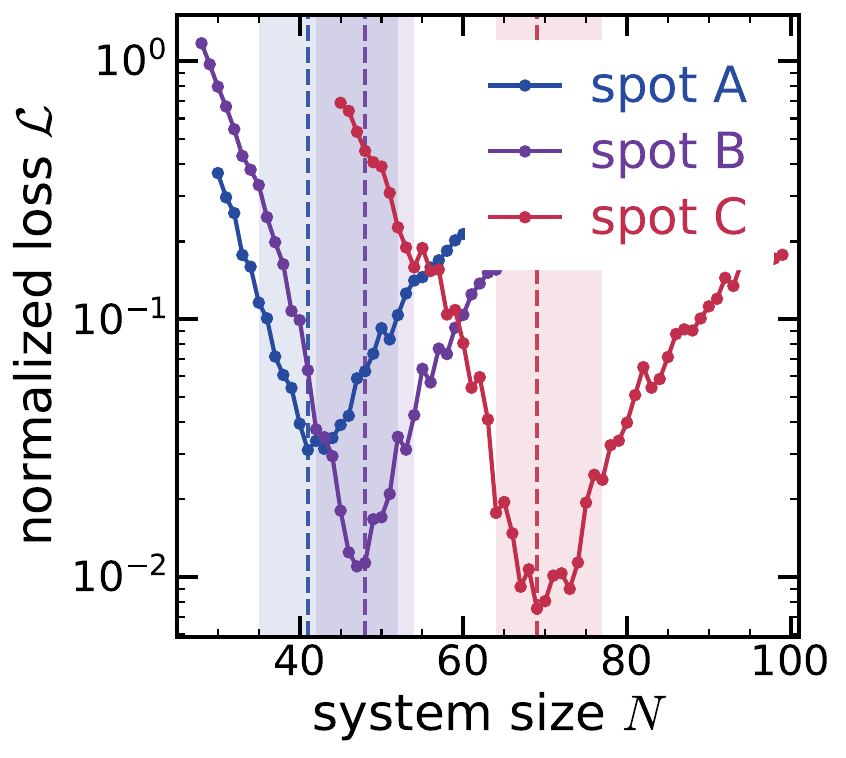}
\caption{%
Normalized fit loss $\mathcal{L}$ [Eq.~\eqref{eq:sim-loss}] as a function of
the system size $N$ for spots A (blue), B (purple), and C (red). Dashed lines
mark the adopted values, and the shaded regions the accepted range
$\mathcal{L} \le 4\mathcal{L}_{\rm min}$, which sets the uncertainty quoted
on $N$.
}
\label{fig:supp_readoutsim}
\end{figure}

\subsection{Extraction of the system size from XY-8}
\label{sec:XY-8-size}

The system sizes quoted in Table~\ref{tab:twisting-parameters} are obtained in
two independent ways. The first, described in the preceding part, fits
the measured variance with $N$ as the only free parameter. Here we describe the
second method, which uses the NV--NV interactions. The number of NVs in a confocal volume sets the typical dipolar coupling strength between them. This coupling strength in turn sets the timescale on which the interactions destroy the coherence of the ensemble. We
measure this timescale by preparing the NV ensemble in the $\ket{x}_e^{\otimes N}$ product state and letting it time-evolve under an XY-8 dynamical-decoupling sequence, where
the spins are decoupled from the surrounding paramagnetic bath.

The relation between the decoherence timescale and the density follows from a simple
scaling argument. Because the NVs are confined to a delta-doped layer much
thinner than their typical separation, the ensemble is effectively
two-dimensional. In this system, the scaling is:
\begin{equation}
  T_2(\rho) = T_2(\rho_0)\left(\frac{\rho_0}{\rho}\right)^{3/2},
  \label{eq:powerlaw}
\end{equation}
where $\rho$ denotes the areal density of the (single, driven orientation
group of) NVs. We fix the exponent to this two-dimensional value
throughout. 
It remains to fix the prefactor in Eq.~\eqref{eq:powerlaw}, which we calibrate against the cluster discrete-truncated-Wigner (DTWA) simulations published alongside Ref.~\suppcite{SMref:Wu2025}. In those simulations, $N=80$ spins are placed at random positions in a slab of thickness $6.8\,\mathrm{nm}$ (uniform in $z$, Poisson-distributed laterally). NVs evolve under the dipolar XXZ Hamiltonian, and the signal is averaged over $50$ positional-disorder realizations, for areal densities $\rho \in \{6,7,8,9,10\}\,\mathrm{ppm\cdot nm}$.
Fitting each disorder-averaged coherence to the two-dimensional positional-disorder form~\suppcite{SMref:Davis2023}
\begin{equation}
  C(t) = A\, e^{-(t/T_2)^{2/3}}
  \label{eq:stretch}
\end{equation}
gives the values in Table~\ref{tab:dtwa_T2}. Then, a fit of these five points to Eq.~\eqref{eq:powerlaw} yields  $T_2(\rho_0) = 12.4(11)\,\mu$s at $\rho_0 = 8\,\mathrm{ppm\cdot nm}$
[Fig.~\ref{fig:supp_density}].

\begin{table}[h]
\centering
\begin{tabular}{c|ccccc}
\hline\hline
$\rho$ (ppm$\cdot$nm) & 6 & 7 & 8 & 9 & 10 \\
$T_2$ ($\mu$s) & 16.5 & 15.0 & 12.6 & 11.0 & 9.6 \\
\hline\hline
\end{tabular}
\caption{Decoherence timescales under the NV--NV dipolar interactions extracted from the cluster-DTWA simulations of
Ref.~\protect\suppcite{SMref:Wu2025} by fitting Eq.~\eqref{eq:stretch}.}
\label{tab:dtwa_T2}
\end{table}

Fitting the measured XY-8 decoherence of spots A (interpulse spacing
$\tau_p = 100\,\mathrm{ns}$), B ($\tau_p = 300\,\mathrm{ns}$),
C ($\tau_p = 200\,\mathrm{ns}$) and D
($\tau_p = 300\,\mathrm{ns}$) to Eq.~\eqref{eq:stretch}, we obtain
$T_2^{\mathrm{A}} = 53(3)\,\mu$s, $T_2^{\mathrm{B}} = 160(17)\,\mu$s,
$T_2^{\mathrm{C}} = 44(4)\,\mu$s and
$T_2^{\mathrm{D}} = 10.0(7)\,\mu$s. Here, in the experiment, the interpulse spacing of the XY-8 sequence on each spot is chosen so that $T_2$ is maximized.
Eq.~\eqref{eq:powerlaw} then gives
\begin{equation}
  \rho_{\mathrm{A}} = 3.0(2)\,\mathrm{ppm\cdot nm},
  \qquad
  \rho_{\mathrm{B}} = 1.5(2)\,\mathrm{ppm\cdot nm},
  \qquad
  \rho_{\mathrm{C}} = 3.5(3)\,\mathrm{ppm\cdot nm},
  \qquad
  \rho_{\mathrm{D}} = 9.2(11)\,\mathrm{ppm\cdot nm}.
\end{equation}

Finally, we convert the density into a spin number. Each NV contributes to the
measured signal with a weight $W(\mathbf r)$ that depends on its position
$\mathbf r$ relative to the center of the focus, given by the product of the
green excitation and red collection profiles. Modeling both as
diffraction-limited Gaussians at $532$ and $637\,\mathrm{nm}$ for
$\mathrm{NA} = 0.95$, 
the effective spin number weighted in
this way is $N_{\mathrm{eff}} = \rho_{\mathrm{2D}}\,\pi w_{\mathrm{eff}}^2$, where $w_{\mathrm{eff}} = 186\,\mathrm{nm}$ is the effective $1/e$ radius.
Allowing for the focus to be slightly larger than the diffraction limit (we
take the spot area to be $30\%$ above this value with a $30\%$ uncertainty), we obtain
\begin{equation}
  N_{\mathrm{eff}}^{\mathrm{A}} = 75(23), \qquad
  N_{\mathrm{eff}}^{\mathrm{B}} = 36(12), \qquad
  N_{\mathrm{eff}}^{\mathrm{C}} = 86(27), \qquad
  N_{\mathrm{eff}}^{\mathrm{D}} = 229(74),
\end{equation}
corresponding to the values quoted in Table~\ref{tab:twisting-parameters}.

\begin{figure}[t]
\centering
\includegraphics[width=0.45\linewidth]{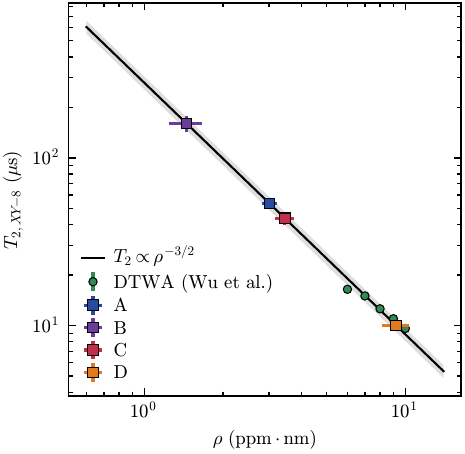}
\caption{
Coherence time $T_{2,XY\text{-}8}$ versus the areal density $\rho$ of the driven
NV group, extracted from the cluster-DTWA simulations of
Ref.~\protect\suppcite{SMref:Wu2025} (green circles). The black line is
Eq.~\eqref{eq:powerlaw} with the exponent fixed to the two-dimensional value
$3/2$ and the prefactor $T_2(\rho_0)$ fitted to the five points; the gray band is
its uncertainty. Squares mark the measured coherence times of spots A--D,
and the densities inferred from them.
}
\label{fig:supp_density}
\end{figure}

\subsection{Readout noise improvement by repetitive readout}
\label{sec:sigma-r-enhancement}
To quantify the improvement afforded by repetitive readout, we compare its readout noise $\sigma_R^{\rm rep}$ with that of the first, single conventional readout in the repetitive-readout cycle,
\begin{equation}
  \sigma_R^{\rm conv} = \sqrt{1 + \frac{2(a_1 + b_1)}{(a_1 - b_1)^2}},
  \label{eq:sigma-r-conv}
\end{equation}
where $a_1$ and $b_1$ are the mean photon counts of the states $\ket{0}_e$ and $\ket{-1}_e$ of a single NV~\cite{SMref:Hopper2018}. In the experiment, we obtain $a_1$ and $b_1$ by dividing the measured photon count of the ensemble by the system size $N$, which is extracted from the fit to the repetitive readout's $\sigma_{\rm res}^2$ data. Furthermore, the readout duration is swept to minimize $\sigma_{R}^{\rm conv}$. Table~\ref{tab:sigma_r_comparison} compares the two readout schemes for the four spots A, B, C and D in the main text.

\begin{table}[h]
\centering
\begin{tabular}{c|ccc}
\hline\hline
Spot & $\sigma_R^{\rm conv}$ & $\sigma_R^{\rm rep}$ & Enhancement \\
\hline
A & 18.2 & 6.8 & 2.7\\
B & 17.9 & 2.7 & 6.6\\
C & 9.5 & 3.1 & 3.1\\
D & 87 & 38 & 2.3\\
\hline\hline
\end{tabular}
\caption{Readout noise of a single conventional readout, $\sigma_R^{\rm conv}$, compared with that of the repetitive readout, $\sigma_R^{\rm rep}$, for the four spots. The enhancement is the ratio $\sigma_R^{\rm conv}/\sigma_R^{\rm rep}$. The values of $\sigma_R^{\rm rep}$ are those reported in Table~\ref{tab:twisting-parameters} of the main text.}
\label{tab:sigma_r_comparison}
\end{table}

\subsection{Optimization of readout laser parameters}
\label{sec:laser-power-opt}

The performance of the repetitive readout depends on how we operate the green laser. We can tune three parameters, namely the laser power $P$, the electronic-spin initialization time $T_{\mathrm{init}, e}$, and the duration of each readout $T_{r}$. These parameters are not independent, and optimizing the readout noise $\sigma_R$ requires balancing several competing timescales.

First, the two timescales $T_{\mathrm{init}, e}$ and $T_{r}$ depend on the laser power $P$. For a fixed $P$, there is a timescale over which the electronic spin is repolarized to $\ket{0}_e$ between shots. Then, in the repetitive readout, one needs to set the initialization time $T_{\mathrm{init}, e}$ comparable to this electronic-spin repolarization timescale. Moreover, the laser power sets an optimal readout duration that mainimizes $\sigma_R$ of a single conventional readout: A longer pulse collects more photons, but once the electronic spin has been initialized, the additional photons only add photon shot noise. We find that in our sample, it is advantageous to stop the electronic-spin initialization after roughly the optimal readout duration, even though this leaves the electronic spin imperfectly initialized, because it shortens each cycle and thereby increases the repetition rate. We emphasize that these conclusions depend on the specific sample.

Second, the laser power enters through a trade-off with the nuclear memory. As discussed in the main text, a lower laser power excites the electron less often and therefore lengthens the nuclear depolarization time $T_{1n}$. At the same time, it also lengthens the readout and initialization times. The optimal operating point is the power that best balances a long $T_{1n}$, which allows more repetitions, against the time overhead of each shot. For our sample, we find that the lowest accessible laser power of $10~\si{\mu W}$ gives the best readout noise, because the resulting long nuclear memory outweighs the longer per-shot overhead.

\subsection{Readout noise cannot generate anisotropy of \texorpdfstring{$\sigma^2_{\rm res}$}{sigma_res^2}}
\label{sec:sm_readout_isotropy}

The repetitive readout chain adds noise to the intrinsic quantum projection noise, and the magnitude of that added noise depends on the projected state. It is therefore natural to ask whether the readout could
also generate the $\phi$-dependence of $\sigma^2_{\rm res}$ reported in Fig.~\ref{fig:fig-3} of the main text. Here we show that it cannot. Because every readout imperfection acts independently on each spin after its projection,
it enters $\sigma^2_{\rm res}$ only through the first moment $\langle S_\phi^{\rm tot} \rangle = \langle S_z^{\rm tot}\,\cos\phi + S_y^{\rm tot}\,\sin\phi \rangle$ and through a $\phi$-independent rescaling of the true spin variance. Since the twisting dynamics keep the mean spin pinned along $\hat{x}$, $\langle S_\phi^{\rm tot} \rangle$ vanishes and all readout-induced noise is isotropic.

Let $S_\phi^{\rm tot}$ denote the measured collective spin component as in the main text, and let $s_i \in \{0,1\}$ be the projection outcome of spin $i$ along $S_\phi^{\rm tot}$. Conditioned on the outcomes $\{s_i\}$, spin $i$ contributes to the weighted photon signal $n$ with mean $\mu_{s_i}$ and variance $v_{s_i}$, independently of the other spins. Here, the parameters $\mu_{0,1}$ and $v_{0,1}$ absorb the full readout chain (imperfect CNOT gates, incomplete electronic repolarization, laser-induced nuclear depolarization, and photon shot noise). Let $N_1 = \sum_i s_i$ denote the number of spins projected onto $s_i = 1$. The conditional mean and variance of the total photon signal $n$ are then
\begin{align}
  \mathbb{E}\!\left[n \,\middle|\, \{s_i\}\right]
    &= N_1 \mu_1 + (N - N_1)\,\mu_0 , \\
  \mathrm{Var}\!\left[n \,\middle|\, \{s_i\}\right]
    &= N_1 v_1 + (N - N_1)\,v_0 ,
\end{align}
so that the total variance is
\begin{equation}
  \mathrm{Var}(n)
  = \underbrace{N\!\left[p\,v_1 + (1-p)\,v_0\right]}_{\text{readout}}
  + \underbrace{(\mu_1 - \mu_0)^2\,\mathrm{Var}(N_1)}_{\text{projection}} ,
  \qquad
  p \equiv \frac{\langle N_1 \rangle}{N}
    = \frac{1}{2}\!\left(1 - \frac{2\langle S_\phi^{\rm tot} \rangle}{N}\right).
  \label{eq:total_variance}
\end{equation}
The first term contains all readout-induced noise, and it depends on the state only through $\langle S_\phi^{\rm tot} \rangle$.
Under twisting generated by the XXZ interaction, the mean spin remains along $\hat{x}$, orthogonal to the
$yz$ measurement plane, so $\langle S_\phi^{\rm tot} \rangle = 0$ for all $\phi$ and all evolution times $t$. Equation~\eqref{eq:total_variance} therefore reduces to
\begin{equation}
  \sigma^2_{\rm res}(t,\phi)
  = \sigma^2_{\rm ro}(t) + C^2\,\mathrm{Var}\!\left[S_\phi^{\rm tot}(t)\right],
\end{equation}
with an isotropic offset $\sigma^2_{\rm ro}$ and a $\phi$-independent
contrast $C$; any measured anisotropy therefore witnesses the spin
variance alone.

Two conditions delimit this argument, and both are satisfied in the experiment. First, a residual tilt of the mean spin out of $\hat{x}$ would give $\langle S_\phi^{\rm tot} \rangle = \epsilon \cos(\phi - \phi_0)$ and hence a small, order $\epsilon$ anisotropic contribution. We verify directly that $\langle S_\phi^{\rm tot} \rangle$ is flat and consistent with zero (within 1\% of the Rabi contrast) at every $t$. 
Second, the readout must act independently on each spin. This condition could be
violated by common-mode mechanisms such as laser intensity fluctuations. Any such contribution to the readout noise is nonetheless isotropic as well,
and the differential subtraction further removes residual common-mode drifts in the experiment. The data confirm this independently: no anisotropy of the residual noise $\sigma^2_{\rm res}$ is observed in the time evolution under DROID-60.

\section{Estimation of sensitivity enhancement}

The suppression of classical readout noise demonstrated in the main text translates into a quantitative metrological advantage. The relevant figure of merit is the magnetometric sensitivity,
\begin{equation}
  \eta = \frac{\sigma_R}{\gamma_{\mathrm{e}}\, T_{\mathrm{sense}}\, \sqrt{N}} \sqrt{T_{\mathrm{tot}}},\label{eq:sensitivity}
\end{equation}
where $\gamma_{\mathrm{e}}$ is the electron gyromagnetic ratio, $T_{\mathrm{sense}}$ is the sensing duration, and $T_{\mathrm{tot}} = T_{\mathrm{init}} + T_{\mathrm{sense}} + T_R$ is the total cycle time, which includes the state-preparation time $T_{\mathrm{init}}$ and the repetitive-readout time $T_R$. As the number of repetitions $k$ increases, the readout noise $\sigma_R(k)$ improves. However, minimizing $\sigma_R$ does not by itself guarantee an improved sensitivity, because repetitive readout imposes a time overhead $T_R$ roughly proportional to $k$ that eventually inflates $T_{\mathrm{tot}}$. Achieving a net sensitivity gain therefore requires optimizing the parameters in Eq.~\eqref{eq:sensitivity}. In particular, the sensing time $T_{\mathrm{sense}}$, which is typically bounded by the spin coherence time, must be long compared to $T_R$, so that the repetitive readout does not dominate the total experimental time overhead.

Figure~\ref{fig:sensitivity} shows the enhancement of the volume-normalized sensitivity $\eta_V$ for spot B of our sample, as a function of the repetition number $k$ and the sensing time $T_{\mathrm{sense}}$. For sufficiently long sensing times ($\gtrsim 1~\si{ms}$), the repetitive readout yields a net sensitivity gain. Here, $N$ is obtained from the fit to the variance data described in the main text, and the sensing volume is estimated from the confocal spot size in the lateral directions and from the delta-doped layer thickness in the vertical direction, as laid out earlier in this Supplemental Material.

\begin{figure}[htbp]
\centering
\includegraphics[width=0.6\linewidth]{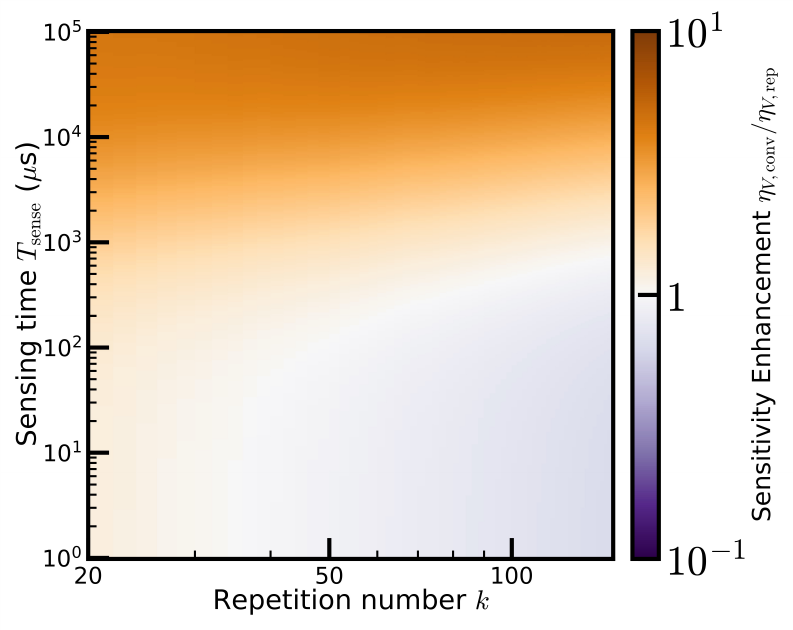}
\caption{Expected enhancement of the volume-normalized magnetometric sensitivity $\eta_V$ from conventional to repetitive readout, for spot B of our sample, as a function of the repetition number $k$ and the sensing time $T_{\mathrm{sense}}$. When the sensing time is sufficiently long compared to the readout overhead ($\gtrsim 1~\si{ms}$), the repetitive readout yields a net sensitivity gain.}
\label{fig:sensitivity}
\end{figure}

\phantomsection
\addcontentsline{stoc}{section}{References}
\makeatletter
\let\supplement@bibliographyaddcontentsline\addcontentsline
\renewcommand{\addcontentsline}[3]{%
  \def\supplement@requestedfile{#1}%
  \ifx\supplement@requestedfile\supplement@tocname
  \else
    \supplement@bibliographyaddcontentsline{#1}{#2}{#3}%
  \fi
}
\makeatother
\supplementbibliography
\endgroup
\supplementfinish

\end{document}